\newif\ifabstract
\abstracttrue
 \abstractfalse 
\newif\iffull
\ifabstract \fullfalse \else \fulltrue \fi

\documentclass[11pt]{article}
\usepackage{amsfonts}
\usepackage{amssymb}
\usepackage{amstext}
\usepackage{amsmath}
\usepackage{xspace}
\usepackage{theorem}
\usepackage{graphicx}
\usepackage{url}
\usepackage{graphics}
\usepackage{colordvi}
\usepackage{colordvi}
\usepackage{subfigure}

\usepackage{natbib}
\usepackage{wasysym}

\usepackage{subeqnarray}
\usepackage{xcolor}
\usepackage{moreverb} 
\usepackage{textcomp} 
\usepackage{epsfig}
\usepackage{color} 
\usepackage{upgreek}
\usepackage{bbm}

\usepackage{dcolumn}
\usepackage{bm}
\newcommand{\overbar}[1]{\mkern 1.5mu\overline{\mkern-1.5mu#1\mkern-1.5mu}\mkern 1.5mu}

\advance \oddsidemargin by -0.8in
\newcommand{\myparskip}{3pt}
\parskip \myparskip

\newsavebox{\astrutbox}
\sbox{\astrutbox}{\rule[-5pt]{0pt}{20pt}}

\begin{document}

\title{Limiting regimes of turbulent horizontal convection. Part II : Large Prandtl numbers}

\author{Pierre-Yves Passaggia\thanks{University of Orl\'eans, INSA-CVL, PRISME, EA 4229, 45072, Orl\'eans, France, Email: pierre-yves.passaggia@univ-orleans.fr {\tt pierre-yves.passaggia@univ-orleans.fr}. Also affiliated: Carolina Center for Interdisciplinary Applied Mathematics, Dept. of Mathematics, University of North Carolina, Chapel Hill, NC 27599, USA}
\and Nadia F. Cohen\thanks{Department of Marine Sciences, University of North Carolina, Chapel Hill, NC 27599, USA. Email: {\tt nadiafc@live.unc.edu}.}
\and Alberto Scotti\thanks{Department of Marine Sciences, University of North Carolina, Chapel Hill, NC 27599, USA. Email: {\tt ascotti@unc.edu}.} 
\and Brian L. White\thanks{Department of Marine Sciences, University of North Carolina, Chapel Hill, NC 27599, USA. Email: {\tt bwhite@unc.edu}.}}

\maketitle

\thispagestyle{empty}

\begin{abstract}
Horizontal Convection (HC) at large Rayleigh and Prandtl numbers, is studied experimentally in a regime up to seven orders of magnitude larger in terms of Rayleigh numbers than previously achieved. To reach Rayleigh up to $10^{17}$, the horizontal density gradient is generated using differential solutal convection by a differential input of salt and fresh water controlled by diffusion in a novel experiment where the zero-net mass flux of water is ensured through permeable membranes. This setup allows for measuring accurately the Nusselt number in solutal convection by carefully controlling the amount of salt water exchanged through the membranes. Combined measurements of density and velocity across more than five orders of magnitude in Rayleigh numbers show that the flow transitions from the Beardsley \& Festa \cite{beardsley1972numerical,ShishkinaW16} regime to the Chiu-Webster {\it et al.}\cite{chiu2008very} regime and frames the present results within the scope of Shishkina {\it et al.}\cite{ShishkinaGL16}, and the companion paper\cite{Passaggia2019LimitigA} theory. In particular, we show that even for large Prandtl numbers, the circulation eventually clusters underneath the forcing horizontal boundary leaving a stratified core without motion.
Finally, previous regime diagrams\cite{hughes2008horizontal,ShishkinaGL16} are extended combining the present results at high Prandtl numbers, the results at low Prandtl numbers of the companion paper, together with previous results from the literature. This work sets a new picture of the transition landscape of horizontal convection across six orders of magnitude in Prandtl number and sixteen orders of magnitude in Rayleigh numbers. 
\end{abstract}



%

\maketitle


\section{Introduction}\label{sec:Intro}
Experimental work in horizontal convection has attracted little attention compared to Rayleigh-B\'enard convection. 
 Despite the analysis of Jeffreys\cite{jeffreys1925fluid}, who showed from fundamental thermodynamics that if there was a differential buoyancy gradient along a constant geopotential height, a residual circulation has to exist, Sandstr{\"o}m's \cite{Sandstrom08} inference that experiments in horizontal convection were to result in a very shallow surface circulation detached from a stratified interior at rest dominated the thinking on the subject for a long time   \cite{sverdrup1942oceans,defant1961physical}. This assumption was later questioned by Rossby\cite{Rossby65} using a laboratory experiment. Rossby showed that horizontal convection may indeed lead to a non-negligible overturning flow with a scaling analysis which provided the first insights that despite the small convective intensity of HC when compared to the B\'enard problem, HC could still produce a substantial residual circulation and be relevant for geophysical applications.\\

A key difficulty in designing laboratory experiments in horizontal convection using heat as stratifying agent is the need to prevent buoyancy gain or losses along surfaces other than the horizontal surface where forcing is applied. Wang \& Huang\cite{wang2005experimental} used nearly complete vacuum in a rectangular container to ensure an insulating boundary, while large slabs of Styrofoam were used in the experiment of Mullarney {\it et al.}\cite{mullarney2004convection}.  Both authors identified a regime transition at $\rm{Ra}\approx 10^{10}$ for differential heating in water.
Each experiment provided similar scaling laws. However, one important difference could be found between experiments: the plume observed in the experiment of Wang \& Huang\cite{wang2005experimental} did not fully reach the bottom while the other experiment showed the contrary. While this may be attributed to the insulating boundary, Gayen {\it et al.}\cite{Gayen14} performed direct numerical simulations of the setup of Mullarney {\it et al.} and recovered a flow very similar to what was observed in the experiment. These experiments thus raised the question of the role of the aspect ratio of the cavity on the flow. Working with differential heating is very appealing at first since the viscosity of water can be increased (hence the Prandtl number) using for instance glycerol. However this method simultaneously decreases the Rayleigh number as in Rossby's original experiment which prevents to reach high Rayleigh numbers in the lab.\\

Large-Prandtl and large-Rayleigh numbers regimes have several important applications, ranging from mantle convection to industrial applications such as glass furnaces \cite{gramberg2007convection,chiu2008very}.
Such regimes were theorized by Gramberg {\it  et al.} \cite{gramberg2007convection}, who assumed that the return flow is distributed over the depth of the shallow layer, which makes the thin light layer move at a uniform velocity to leading order. However these findings were later questioned by Chiu-Webster {\it et al.}\cite{chiu2008very} who showed that a laminar regime exists where only the densest fluid in the stratified boundary layer penetrates through the full box depth, the remainder returning at shallow depth in a horizontal intrusion immediately adjacent to the boundary layer. In this case, diffusion between the interior and the relatively weak full-depth plume is crucial for both, removing the density anomaly in the plume fluid and maintaining a stratification in the box interior. Very recently, Ramme \& Hansen \cite{ramme2019transition} investigated a similar regime but for higher Rayleigh numbers using two-dimensional numerical simulations. They report a transition to a steeper scaling  than previously reported in \cite{chiu2008very}. They report that the distribution of the forcing only has a small impact on the dynamics and similarly for the effect of shear-free or no-slip boundary conditions. They also noticed that the onset of instability introduces a steeper scaling for the Nusselt and P\'eclet numbers at Rayleigh numbers between $10^8$ and $10^9$ which they conjectured to be associated with the transition observed in Shishkina \& Wagner \cite{ShishkinaW16}. However, it should be recalled that the transition observed in ref. \cite{ShishkinaW16} was not induced by the onset of instabilities. It should also be noted that local steeper scaling than the one observed in Rossby's theory, as for instance reported in ref.\cite{Gayen14}, is associated with stability transitions. In addition, the effect of the aspect ratio remains to be examined at large Prandtl numbers since in Shishkina \& Wagner theory \cite{ShishkinaW16}, the circulation has to span the entire depth of domain, hence requiring for instance domains with large aspect ratios.

However to this date, no experiments nor numerical simulations at large Rayleigh and Prandtl numbers have yet been performed. Thus it is not clear whether the circulation transitions to an intrusion-type regime or whether other transitions may be explored as the Rayleigh number increases towards geophysical applications. 
A flow exhibiting a shallow circulation close to the differential buoyancy forcing is known as the intrusion regime and we leverage on the results of Chiu-Webster {\it et al.}\cite{chiu2008very} to analyze the present experimental results.\\

Although it is customary to refer to the ratio of viscosity to diffusivity of a solute as the Schmidt $({\rm Sc})$ number, for consistency with the previous paper and with most of the HC literature, here we call such a ratio the Prandtl number. To achieve high Prandtl and Rayleigh number flows we us a combination of long tanks with large aspect ratios and a weakly diffusive stratifying agent. Diffusion of salt ions is a simple and effective alternative to heat, in order to modify the buoyancy of a fluid. Moreover, solid boundaries naturally act as no-flux boundaries. To impose the boundary condition at the surface,  we use large sheets of semi-permeable 
membranes stretched over rectangular tanks of different sizes, which could allow for reaching Rayleigh numbers up to nearly $10^{18}$. As an added benefit, using double emission laser induced fluorescence makes it possible to measure the buoyancy field\cite{PassaggiaHSW17}. Using this setup,
our aim is to complete the map from the companion paper \cite{Passaggia2019LimitigA} and extend the regime diagram of HC to large values of the Prandtl number.

Griffiths \& Gayen \cite{griffiths2015turbulent} considered the problem of horizontal convection forced by spatially periodic forcing. Their results showed that for high Rayleigh number and small aspect ratios, the core would indeed fill with dense fluid and maintain a stratified interior. Their forcing, localized on a length scale smaller than the depth of the domain, and with variation in both horizontal directions show turbulence throughout the domain. Associated experiments were performed by Rosevear {\it et al.} \cite{rosevear2017turbulent} where they observed that the Nusselt number (a non-dimensional measure of the buoyancy flux) had a steeper scaling with respect to the Rayleigh number than the (laminar) Rossby scaling\cite{Rossby65} or the entrainment regime\cite{hughes2008horizontal} and the intrusion regime\cite{chiu2008very}. Here we revisit their experiments replacing porous brass plates with permeable membranes\cite{krishnamurti2003double}, which  allow for accurate measurements of the Nusselt number, as previously suggested in the experiment of Mullarney {\it et al.}\cite{mullarney2004convection}. We further improve the method by providing evidence that steady states are reached for each experiments in a self-validating experimental setup.

More recently, Matusik \& LLewelyn-Smith\cite{matusik2019response} analyzed the response of surface buoyancy flux‑driven convection to localized mechanical forcing where salt and fresh water fluxes input directly into the tank with pumps whereas the excess of water is set to leave as an over flow. This setup has the advantage to drive a localized surface forcing but can not be considered as a closed system solely driven by buoyancy. Whitehead \& Wang\cite{whitehead2008laboratory} and  Stewart {\it et al.} \cite{stewart2012role} experiments are also worth mentioning in this context. They used mechanical stirring in the interior of the tank to analyze the relation between mixing in the interior and the response on the circulation.\\

In this study, we report experimental results on how the Nusselt number ($\rm{Nu}$) and the boundary layers' thickness ($\lambda$), the streamfunction ($\Psi$) and approximate values of the Reynolds number at the center of the domain depend on the Rayleigh number ($\rm{Ra}$), the flux Rayleigh number (${\rm Ra}_f$) and the Prandtl number ($\rm{Pr}$) in HC at high $\rm{Pr}$ (i.e. equivalently $\rm{Sc}$) for values characteristic of solutal convection in salt where $\rm{Pr}\equiv{\rm Sc}\approx 610$.\cite{harned1954diffusion} The results are in agreement with the scaling power laws recently derived by Shishkina {\it et al.} \cite{ShishkinaGL16} based on the original work of Grossmann \& Lohse \cite{GL00} (GL), the theory of Hughes \& Griffiths \cite{hughes2008horizontal}, previous experiments (\cite{miller1968thermally,mullarney2004convection}), and previous numerical simulations \cite{beardsley1972numerical,Rossby98,ilicak2012simulations}.
Our experiments cover the laminar regime $I^+_l$ (see ref.\cite{shishkina2017scaling,ramme2019transition,reiter2020classical}), and high-$\rm{Pr}$-high-$\rm{Ra}$ laminar regime $I_u$ (see ref.\cite{chiu2008very}).
The results are discussed and mapped within a landscape regrouping, to the best of our knowledge, all simulations and experiments perform to this date in HC. We show that the $(\rm{Ra,Pr})$ landscape first proposed in the review of Hughes \& Griffiths (see ref.\cite{hughes2008horizontal}) fits within the theoretical prediction Shishkina {\it et al.} \cite{ShishkinaGL16} and blends all known regimes of horizontal convection while recovering all these regimes within a single global picture of HC.\\

The remaining of the paper is organized as follows: the flow setup and experimental apparatus is presented in \S\;II. Previous theoretical scaling derived for large Prandtl number flows laws are then presented in \S\;III and tested against experimental results in \S\;IV. The results are from the companion paper at intermediate- and low-Prandtl numbers are discussed in \S\;V while conclusion are drawn in \S\;VI.

\section{Problem description}

We consider here the problem of convection in the Boussinesq limit, where the density difference $\Delta\rho=\rho_{max}-\rho_{min}$ across the horizontal surface is assumed to be a small deviation from the reference density $\rho_{min}$ taken as the fresh water density at room temperature. 
We use a Cartesian coordinate system where the velocity vector is $\mathbf{u}=(u,v,w)^T$, $b=-g(\rho-\rho_{min})/\rho_{min}$ is the buoyancy, $g$ is the acceleration of gravity along the vertical unit vector $\mathbf{e}_z$.
The control parameters are the Prandtl and Rayleigh number, given by 
\begin{equation}
 {\rm Pr}=\frac{\nu}{\kappa} \quad \mbox{ and } \quad {\rm Ra}=\frac{\Delta L^3}{\nu\kappa},   
\end{equation}
where $\nu$ and $\kappa$ are the viscosity and stratifying agent's diffusivity respectively, $L$ is the horizontal length scale of the domain and $\Delta=-g(\rho_{max}-\rho_{min})/\rho_{min}$. The tank is a parallelepiped of aspect ratio $\Gamma=L/H=16.6$ with dimensions $[L,W,H]/L=[1,1/20,1/16.66]$ where $W$ is the width of tank. Here the length $L=[0.5, 1.21, 2, 4.87]$m are reported. The buoyancy flux is imposed at the top surface $z=H$ where $H$ is the height of the tank.\\

\begin{figure}[t]
\hspace{-0mm}\includegraphics[width=\textwidth]{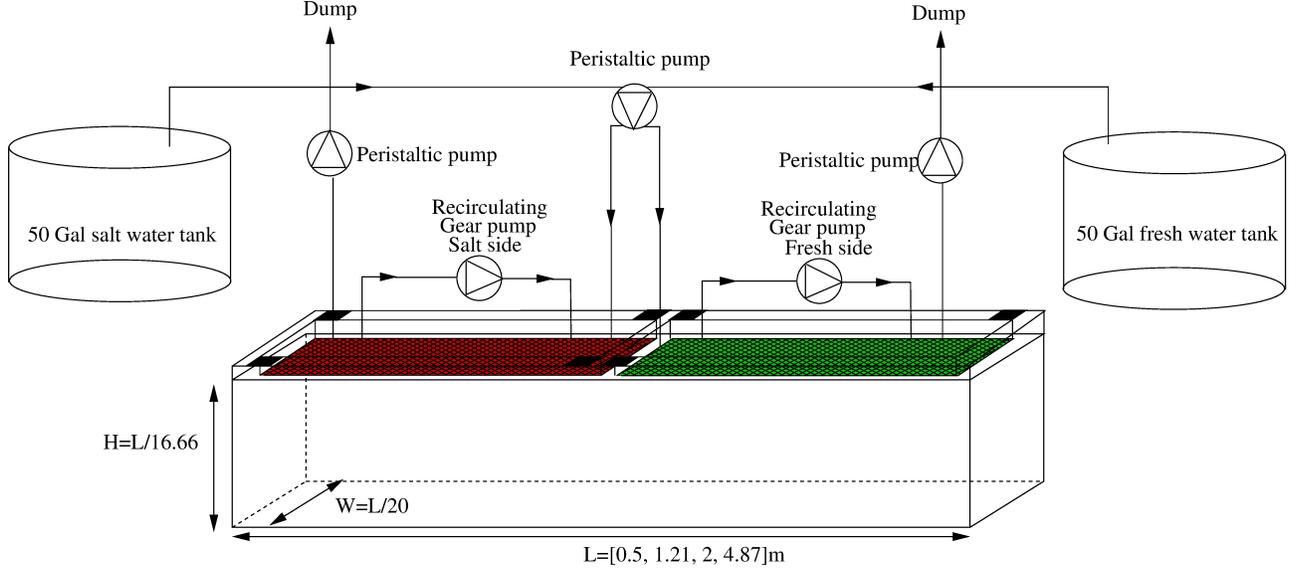}
 \caption{Schematics of the present setup showing the tank with the fresh-water well on the right (green) and salt-water well on the left (red) set over the free surface and constantly stirred to maintain a uniform salinity in the well.}
\label{schematic}

\end{figure}

Distances and flow quantities are non-dimensionalized using the length of the tank $L$ as reference length and the buoyancy difference imposed along the non-isolating horizontal boundary such that
\begin{equation}
\mathbf{x}=\mathbf{x}^*/L, \;\; b=b^*/\Delta , \;\; \mathbf{u}=\mathbf{u}^*/\sqrt{L\Delta}.
\end{equation}

In what follows, we define values to be linked with the control parameters $\rm{Ra}$ and $\rm{Pr}$: the magnitude of the large-scale flow $\Psi_{max}$ is defined as the maximum of the streamfunction, defined as $\Psi=\partial u/\partial z$ and measured at the center of the domain. At the same location, we define $\lambda_u$ and $\lambda_b$, the thickness of the circulation (i.e. the vertical position or height of $\Psi_{max}$) and the thickness of the stratification measured at $({\rho(z)-\rho(z)_{min}})/({\rho(z)_{max}-\rho(z)_{min}})=0.5$. Finally the Nusselt number is expressed as the mass flow rate in and out of the system. The flux Rayleigh number ${\rm Ra}_f$ is defined as the ratio between the amount of salt input and/or output from the tank, normalized by dissipation and writes
\begin{equation}
{\rm Ra}_f=\frac{\dot{Q}g(\rho_{well}-\rho_{tank})/\rho_{tank}L^4}{\nu\kappa^2}
\end{equation}
where $g=9.81$ m/s$^2$ is the acceleration of gravity, $\dot{Q}$ is the volume flow rate of fluid input into the system while the subscripts $_{tank}$ and $_{well}$ define where the measurements of density were acquired (see the following section). The later can be used to obtain the Nusselt number by computing the ratio
\begin{equation}
{\rm Nu}={\rm Ra}_f/{\rm Ra},
\label{eq:NuRafRa}
\end{equation}
which is the ratio between the convective buoyancy flux, normalized with the conductive buoyancy flux imposed at the boundary. Note that a very similar technique was employed in previous experiments \cite{mullarney2004convection,griffiths2013horizontal,rosevear2017turbulent} to compute the Nusselt number in the case where a flux is imposed through the boundary.


\begin{figure}[t]
\hspace{-0mm}\includegraphics[width=\textwidth]{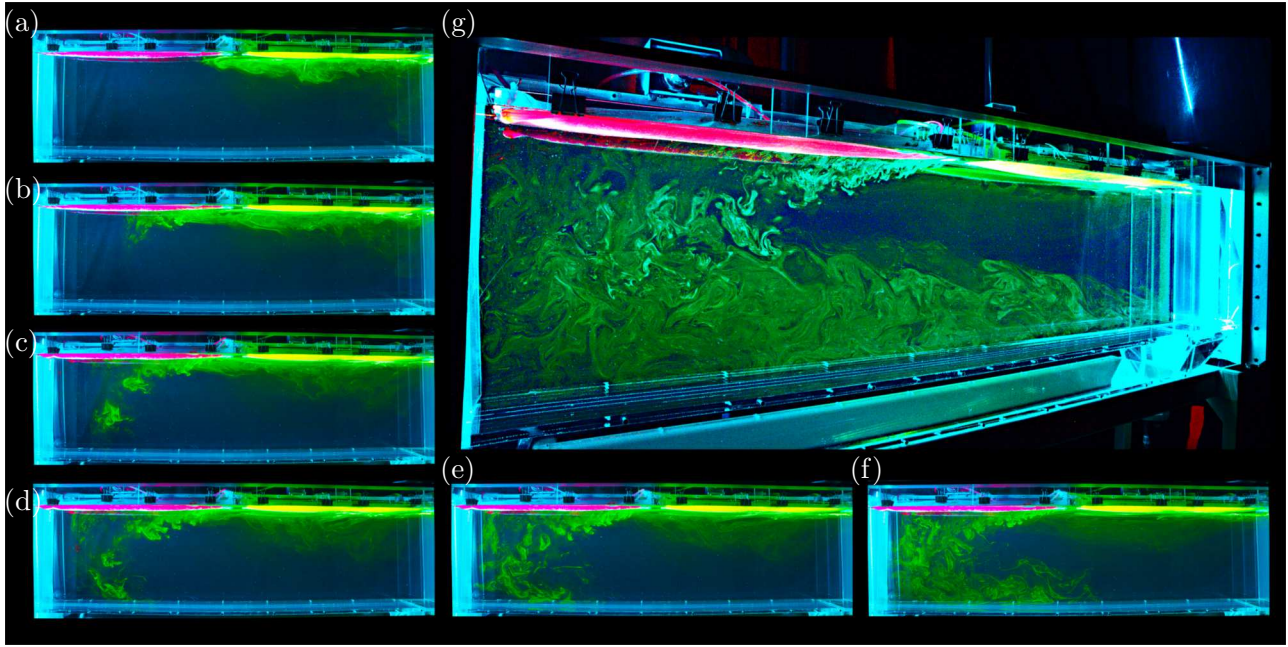}
\put(-485,232){\color{white}(a)}
\put(-485,169){\color{white}(b)}
\put(-485,111){\color{white}(c)}
\put(-485,51){\color{white}(d)}
\put(-320,63){\color{white}(e)}
\put(-165,63){\color{white}(f)}
\put(-320,232){\color{white}(g)}
 \caption{Side view of the tank, illuminated from the left and showing the evolution of fresh fluorescein
dye water released from the right. The temporal evolution of the circulation is described by the transport of
the fluorescein (green/yellow) dye, driven by a solutal horizontal density gradient, creating a turbulent plume
sinking on the left of the tank (a-f). The deep circulation is shown in (g) where the flow has developed and drives a deep but weakly turbulent circulation that eventually upwells to the right of the tank. https://doi.org/10.1103/APS.DFD.2016.GFM.P0028}
\label{DELIF}
\end{figure}


The present series of experiments consist of four geometrically similar acrylic tanks of various lengths $[0.5,1.21,2,4.87]$ m.  The tanks were kept at a constant temperature and covered to prevent convection induced by the building's ventilation. On top of the free surface, Spectrum Labs\texttrademark
Spectra/Por$^{\mbox{\textregistered}}$ 5 Reinforced 12-14 kD, 0.280m wide and up to 15m long permeable membranes were stretched and kept from sagging into the tank, forming two separate shallow wells, one filled with fresh water, the other with salty water.  Each well was stirred continuously using a gear pump to keep uniform densities in the wells. Each well was provided with fresh and salt water respectively at a constant flow rate of $\dot{Q} = [10,40,40,80]$ ml/min using a 600 RPM Cole-Parmer\texttrademark 7523-80 Digital Peristaltic Pump. Note that new tubing was used for each experiment. The wells were fed from two 200-liter tanks, whose capacity was selected so that they could supply even the longest running experiments (a couple month) without interruption.  Density   measurements were recorded using an Anton$^\textsf{\textregistered}$ Paar DM35 densitometer whose calibration was verified to the fourth digit.
Each experiment was illuminated using a laser from the left to the right for the two smaller tanks. In the case of the larger tanks, a laser light sheet was introduced between the two wells and illuminated the tank at mid-distance. At the same location, conductivity measurements were performed using a Conduino\cite{carminati2017conduino} probe to obtain high-resolution profiles of conductivity and therefore density. The Conduino probe was calibrated using the Anton$^\textsf{\textregistered}$ Paar DM35 densitometer. It was also checked that the temperature variation was less than 0.3 $^o$C between the top and the bottom of the tanks, resulting in a 6\% relative buoyancy difference, compared to the buoyancy difference introduced by the salt, in the case of the smallest differential solutal input imposed in the wells with the salt.\\

Planar two-dimensional PIV data were recorded using a Nikon D4$^\textsf{\textregistered}$. We used a continuous green laser-pointer at $532$nm whose beam was expanded through a double concave lens. The camera was equipped with a Nikon$^\textsf{\textregistered}$ AF-S VR Micro-Nikkor 105mm f/2.8G IF-ED lenses and the analysis of the experimental data was performed using the Matlab$^\textsf{\textregistered}$-based PIV software DPIVSoft \citep{meunier2003analysis,passaggia2012transverse} to process the images. The resolution of the PIV was $0.0001$ cm/px in the worst case which allowed for fully resolving the PIV particles. The time between two pictures was varied between 1 and 3 seconds.
The top layer was seeded using Cospheric$^\textsf{\textregistered}$ neutrally buoyant for $\rho=[1.000, 1.02, 1,13 ]$ g/cc mono-disperse polyethylene PIV particles with diameters in the range [40:50] $\mu$m which were wetted before hand with the top layer fluid, mixed in a separate tank, reinjected slowly in the free  surface in between the wells, and left to slowly settle across the layers for a couple hours until they reached their position of neutral buoyancy. The subsequent analysis shows that this time scale is sufficient to get back to a steady state when disturbing the flow.\\

A snapshot of the experiment is shown for illustration in Fig. \ref{DELIF} for $A=4$ (i.e. a separate experiment from the rest of this study) where the early stage of the experiment (i.e. the transient phase of the flow) can be visualized. Dye was released on the right part of the tank, near the stable layer and propagated within the turbulent plume which eventually fills the the tank with heavier fluid. While this series of images seem to resemble to the dynamics shown in previous experiments using heat \cite{mullarney2004convection}. The steady state reached in the case of salt is rather different and the remainder of the manuscript links theoretical arguments to direct observations and measurements of density and velocity profiles in order to determine the nature of the HC flow at large Rayleigh and large Prandtl numbers.\\

\section{Scaling and regimes of horizontal convection at large Prandtl numbers}\label{sec:preregimes}

We begin with reviewing the existing scaling laws derived in the limit of large Prandtl numbers and report the exponents for heat and momentum exchanges in horizontal convection. It is interesting to note that for large Prandtl numbers regimes, the regime diagram in the $(\rm{Ra,Pr})$ plane (i.e. $(\rm{Ra,Sc})$) from the review of Hughes \& Griffiths \cite{hughes2008horizontal} does not agree with the regime diagram suggested by Shishkina {\it et al.} \cite{ShishkinaGL16}. In the next subsection, we review these regimes and point out the subtle difference characterizing each of them. We also leverage on the Paparella \& Young \cite{PaparellaY02} inequality which relates the mean mechanical dissipation of the system with the input of heat through the horizontal boundary to discuss the role of the importance of the aspect ratio of the domain at large Prandtl numbers.

\subsection{Rossby's (1965) laminar regime $I_l$}

Rossby's laminar regime can be obtained, starting from the steady buoyancy boundary layer equation, which is obtained from the Navier-Stokes equations in the Boussinesq limit \cite{Passaggia2019LimitigA} and allows for writing an advection-diffusion balance in the boundary layer (see the companion paper\cite{Passaggia2019LimitigA} for a thorough derivation)
\begin{equation}
u b_x + v b_z = \kappa b_{zz}.
\end{equation}
The dominant terms in this expression reduce to $U\Delta/L = \kappa \Delta/\lambda_b^2$ where $\lambda_b$ is the thickness of the thermal BL, which scales as $\lambda_b/L \sim \rm{Nu}^{-1}$. Combining the above reduces to the well known thermal-laminar BL scaling
\begin{equation}
\rm Nu=Re^{1/2}Pr^{1/2},
\label{Nu}
\end{equation}
and provides a relation tying $\rm{Nu}$, $\rm{Re}$ and $\rm{Pr}$.
Noting that the thickness of the laminar boundary layer scales as $\lambda_u/L \sim Re^{-1/2}$, the scaling for the mean dissipation in the particular case of laminar BL\cite{Landau87} is
\begin{equation}
\overbar{\epsilon_u}\sim\nu\frac{U^2}{\lambda^2_u}\frac{\lambda_u}{L}=\nu^3L^{-4}{\rm Re}^{5/2}.
\label{epsulam}
\end{equation}
Combining (\ref{Nu}), (\ref{epsulam}) and (\ref{epsu1}), one recovers the laminar scaling \cite{Rossby65,Rossby98,Gayen14,ShishkinaGL16}
\begin{subeqnarray}
\rm{Re} &\sim& \rm{Ra}^{2/5}\rm{Pr}^{-4/5}, \\
\rm{Nu} &\sim& \rm{Ra}^{1/5}\rm{Pr}^{1/10}.
\label{lam_scal}
\end{subeqnarray}
By analogy to the notation in the GL theory for RBC \cite{GL00,ShishkinaGL16}, this scaling regime is denoted as $I_l$, where the subscript $l$ stands for low-$\rm{Pr}$ fluids.

\subsection{Paparella \& Young (2002) inequality}

Horizontal and Rayleigh-B\'enard convection are both closed systems driven by the buoyancy flux imposed through their boundaries. Paparella and Young (PY) \cite{PaparellaY02} first performed a spatio-temporal average of the kinetic energy equation leading to the equality
\begin{equation}
\overbar{\epsilon_u} \;=\; \overbar{wb}, 
\label{epsu1}
\end{equation}
where $\overbar{\epsilon_u}$ is the mean kinetic energy dissipation rate $\overbar{\epsilon_u}\equiv\nu\sum_{i,j}(\partial u_j/\partial x_i)^2$. Another condition can be written using the spatio-temporal average of the Navier-Stokes equation\cite{Passaggia2019LimitigA}, and integrating over $z$ leads to
\begin{equation}
\overbar{wb}=\kappa\langle \partial b/\partial z\rangle_{z=H},
\end{equation}
where $\langle\rangle_{z=H}$ denotes the surface and time average at $z=H$. This equality can be recast into an inequality for the buoyancy between the top an the bottom of the domain which writes
\begin{equation}
\overbar{wb}\leq\kappa(\langle b\rangle_{z=H}-\langle b\rangle_{z=0})/H = B(\Gamma/2)\kappa\Delta/L,
\label{PYbound}
\end{equation}
where $1<B<0$ is an arbitrary constant \cite{ShishkinaW16}. The PY inequality thus writes
\begin{equation}
\overbar{\epsilon_u} \;\leq\; B(\Gamma/2)\nu^{3}L^{-4}\rm{Ra}\rm{Pr}^{-2},
\label{epsu2}
\end{equation}
which combined with the original idea of Rossby, opens possibilities for relating the dissipation in the boundary layer or the core with the heat transfer coefficient near the horizontal boundary.\\

One interesting fact is that PY's inequality suggests that as $\rm Ra$ increases while keeping $\rm Pr$ and $\Gamma$ constant, the flow becomes progressively confined under the conducting boundary. This effect is also known as the anti-turbulence theorem and implies that beyond a certain point, the overturning depth scale becomes 
$$h<H,$$
and a zone of stratified fluid nearly at rest will form on the insulating boundary adjacent to the conducting horizontal boundary. In other words, one may recast the PY inequality eq. (\ref{PYbound})
\begin{equation}
\overbar{\epsilon_u} = B/2(L/h)\nu^{3}L^{-4}\rm{Ra}\rm{Pr}^{-2},
\label{epsunew}
\end{equation}
where dissipation is only bounded to the turbulent core of depth $h$ rather than the entire depth $H$. Note that Shishkina {\it et al.}\citep{ShishkinaGL16} refer to $h$ as the large-scale overturning flow in their theory.

This follows Sandstr\"om\cite{Sandstrom16} inference where at large $\rm Ra$ or for high $\rm Pr$, the flow becomes confined to a progressively thinner surface layer and the core becomes a stagnant pool of stratified water\cite{defant1961physical}. Although such regimes were only observed in direct numerical simulations of laminar HC \cite{ilicak2012simulations} at high Pr and theoretically by \cite{chiu2008very} for the same regimes, experiments by Wang \& Huang\cite{wang2005experimental} show the onset of such behaviour at intermediate $\rm Pr$ and intermediate $\rm Ra$.\\

As $\rm{Ra}$ and/or $\rm{Pr}$ increase, the Rossby regime can no longer stand since the thickness of the return flow decreases as $\lambda_u\sim {\rm Ra}^{-1/5}$ for increasing $\rm Ra$ and $\lambda_b\sim {\rm Pr}^{-1/10}$. In other words, the circulation clusters underneath the forcing boundary and leads to two different regimes explained in the next subsections. 

\subsection{Chiu-Webster intrusion regime at High Prandtl numbers}

\begin{figure}[t]
\centering
\hspace{-2mm}\scalebox{1}{\tiny\input{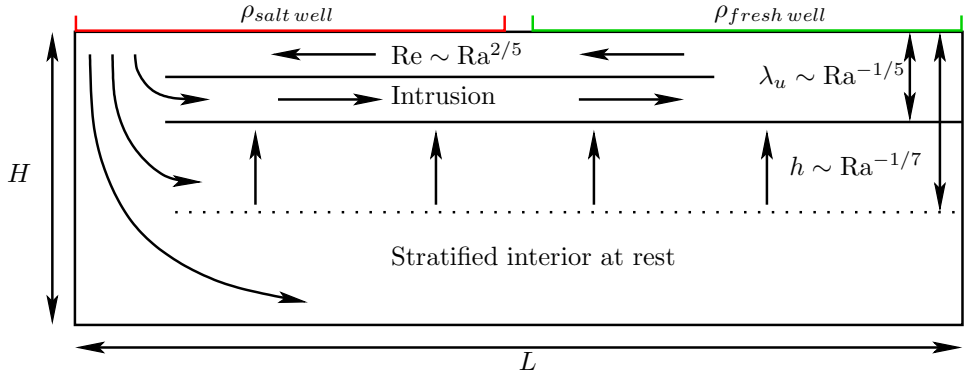}}%
\caption{Schematics showing the intrusion regime of Chiu-Webster {\it et al.}\cite{chiu2008very} and the scaling exponents reported in the present study. Note that the Hughes {\it et al.} regime would correspond to $h=H$ and the intrusion flow scale as $\lambda_b\sim Ra^{-1/5}$ while the Rossby regime would correspond to $\lambda_u \approx H$.}
\label{schematic_intrusion}
\end{figure}

Chiu-Webster, Hinch \& Lister \cite{chiu2008very},  building on the work of Rossby\cite{Rossby65,Rossby98}, recognized that HC is not sensitive to the type of boundary condition (shear-free or no-slip) applied to the velocity field.  In addition, they hypothesized  that the plume dynamics ought to depend on the Prandtl number. While the scaling for the heat and momentum transfer remains essentially the same as Rossby's work, these authors showed that the flow has a more complex structure, characterized by four distinct regions:
\begin{itemize}
    \item A narrow intrusion, clustered underneath the forcing boundary of thickness $\lambda_b\sim\rm{Ra}^{-1/5}$.
    \item A strongly stratified interior where the fluid is at rest and whose thickness scales as $H-h\sim\rm{Ra}^{-1/7}$.
    \item The plume connecting the intrusion layer and the stably stratified interior. 
\end{itemize}

The structure of the flow is depicted in Fig.\ref{schematic_intrusion}.

\subsection{Hughes {\bf{\it et al.}}'s (2007) laminar boundary-layer/turbulent plume regime $II_u$}

Increasing $\rm{Ra}$ and for intermediate $\rm{Pr}$, the kinetic boundary layer becomes progressively thinner while the boundary remains relatively thick in comparison. In this case, it is the thermal boundary layer that drives the dynamics and lead to a turbulent plume and a circulation spanning the entire depth of the domain. This particular case was theorized by Hughes {\it et al.}\cite{Hughes07} with a plume model inside a filling box. Here we recast their model according to the SGL theory (i.e. see the plume model definition eq. (2.15)-(2.20) in  ref.\cite{Hughes07}) and the dissipation in the boundary layer is balanced by the ratio between the thermal and the kinetic boundary layer $\lambda_b/\lambda_u$
which writes
\begin{equation}
\overbar{\epsilon_{u}} \sim\nu\frac{U^2}{\lambda^2_u}\frac{\lambda_u}{L}\frac{\lambda_b}{\lambda_u}= \nu^{3}L^{-4}\rm{Re}^{5/2}\rm{Pr}^{-1/2},
\label{epsuh}
\end{equation}
where the dissipation now scales with the thickness of the thermal layer, not the kinetic BL and is given by $\overbar{\epsilon_{u}}\sim\nu U^2/(\lambda_b L)$. Combining (\ref{Nu}), (\ref{epsu2}) and (\ref{epsuh})
\begin{subeqnarray}
\rm{Re}&\sim&\rm{Ra}^{2/5}\rm{Pr}^{-3/5}, \\
\rm{Nu}&\sim&\rm{Ra}^{1/5}\rm{Pr}^{1/5},
\label{NuRe1/5h}
\end{subeqnarray}
which is denoted as $II_u$ and was first observed in the experiments of Mullarney {\it et al.}\cite{mullarney2004convection} and Wang \& Huang\cite{wang2005experimental}  and later confirmed in the direct numerical simulations of Gayen {\it et al.}\cite{Gayen14}.\\

\subsection{Shishkina \& Wagner (2016) laminar regime $I^*_l$}

At low $\rm{Ra}$ and for large $\rm{Pr}$ and/or large aspect ratio $\Gamma$, the BL thickness $\lambda_u$ saturates and reaches the depth of the domain which gives $\lambda_u=H$ and eq. (\ref{epsu1}) becomes 
\begin{equation}
\overbar{\epsilon_{u}} \sim\nu\frac{U^2}{\lambda^2_u}\frac{\lambda_u}{l}= \nu^{3}L^{-4}\rm{Re}^{2}.
\label{epsul}
\end{equation}
This expression is equivalent to the dissipation of a pressure-driven channel-type flow. Because this regime requires that the boundary layers spans the entire domain, this flow may only be observed for high aspect ratio domains or small Rayleigh numbers which enforces confinement and is the case in the present study.
Combining (\ref{Nu}), (\ref{epsu2}) and (\ref{epsul}), one we obtain the laminar scaling derived in Shishkina \& Wagner \cite{ShishkinaW16}
\begin{subeqnarray}
\rm{Re}&\sim&\rm{Ra}^{1/2}\rm{Pr}^{-1}, \\
\rm{Nu}&\sim&\rm{Ra}^{1/4}\rm{Pr}^{0},
\label{NuRe1/4}
\end{subeqnarray} 
denoted as $I^*_l$ and first observed by Beardsley \& Festa \cite{beardsley1972numerical} in their numerical simulations. It is interesting to note that this scaling bears similarities with the analysis of Gramberg {\it  et al.} \cite{gramberg2007convection} where the return flow takes place along the bottom layer and which may be applicable when the boundary layer spans the entire width of the domain.
Note that Rossby \cite{Rossby98} also observed a steeper scaling than $\rm Nu\sim Ra^{1/5}$ in his numerical simulations for low $\rm Ra$ (see page 248 in \cite{Rossby98}) and similarly in the work of Siggers {\it et al.}\cite{siggers2004bounds} .\\

Note that the Shishkina {\it et al.}\cite{ShishkinaW16} regime $I^*_l$ is expected to occupy the full depth of the domain in Fig.\ref{schematic_intrusion} and we anticipate to see the intrusion regime occur at larger $\rm{Ra}$ than the $I^*_l$ regime.\\

\subsection{The role of the aspect ratio and finite width}

The effect of the domain's aspect ratio was also analyzed by Chiu-Webster {\it et al.} and was reanalyzed by Sheard \& King \cite{sheard2011horizontal} who reached the similar conclusions: for small aspect ratios $A<1$ and for large-enough Rayleigh numbers, the flow follows the $I_u$ regime. For $A\geq2$ and $\rm Pr\gg 1$ the Nusselt number dependence exhibits a slightly steeper scaling and agrees with the conclusions of Shishkina \& Wagner \cite{ShishkinaW16}.
Note that these theories did not take into account the effect of side walls (the influence of non-dimensional width $W/L$) and thus the importance of finite or closed domains. Note that this particular point remains an open question and will not be addressed in the present work.

\subsection{Turbulent regimes at high-Prandtl numbers}

Most of the existing work on HC highlighted  flows driven by laminar-type scaling laws, dominated by the behaviour of the boundary layer, at the exception of an analog of HC Griffiths \& Gayen (2015)\cite{griffiths2015turbulent} and Rosevear {\it et al.}\cite{rosevear2017turbulent}. In a recent study, they considered a spatially periodic forcing at the conducting boundary with a short wavelength compared to the depth of the domain. The companion paper identifies a similar transition but the present study could not achieve the necessary Rayleigh numbers (up to $\rm Ra^{22}$) allowing for observing the transition to a such regime.  \\ 

\section{Experimental results}\label{sec:numerics}

\begin{figure}[t!]
\hspace{-2mm}\scalebox{0.7}{\large\input{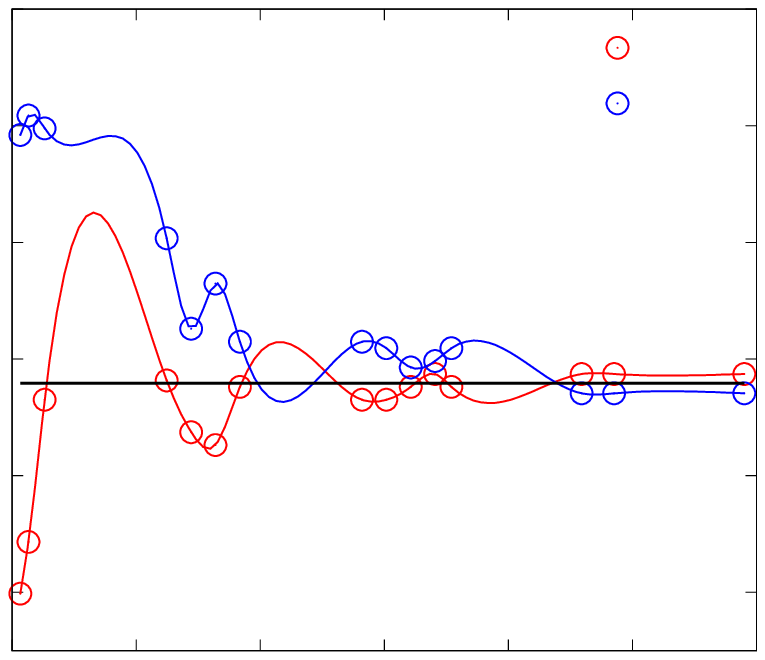}}%
\hspace{-2mm}\scalebox{0.7}{\large\input{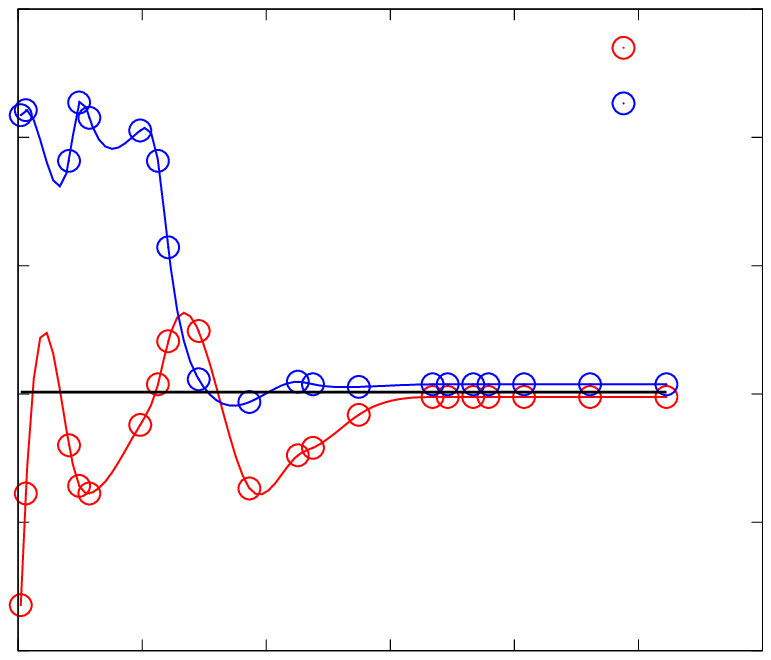}}%
\put(-500,10){(a)}\put(-250,10){(b)}\\
\caption{Temporal evolution of the flux-Rayleigh value in the small tank $L=0.5$m (a) and medium tank $L=2$m (b) over the time rescaled with respect to the diffusion time scale $\tau_d$. The blue curves represents the evolution of the fresh water wells and the red curves represent the evolution of the salt water wells.}
\label{fig:Raf}
\end{figure}

\subsection{Time-scale analysis}

We begin by considering over which time scale a high-Pr HC system will reach a steady state. Clearly, such experiments are possible if the actual time scale is considerable shorter than the purely diffusive time  $\tau_d\sim \kappa^{-1}H^{2}$, which would be close to a year for the largest tank used in our experiments. Luckily for us, convection considerably shortens the transient  by means of stirring the top layer or by entrainment and detrainment in the plume, as seen in Fig. \ref{DELIF} and Fig. \ref{schematic_intrusion}. The amount of buoyancy transported along the horizontal direction over the distance $L$ is given by the streamfunction $\Psi$ and the thickness of the pycnocline which follow the scaling laws mentioned in the above, that is 
\begin{equation}
\Psi_{max}/\kappa \approx c_1 {\rm Ra}_f^{1/6}  \quad \mbox{ and } \quad \lambda_b / L = c_2 {\rm Ra}_f^{-1/6}.
\label{eq:Raf_scaling}
\end{equation}
The prefactors $c_1$ and $c_2$ are obtained from the scaling analysis. In this horizontal convection setup, Griffith {\it et al.}\cite{griffiths2013horizontal} showed that the stable layer acts as a buffer to adjustments or imbalance imposed at the boundaries. In the stably stratified region, the conductive flux is the only mediator to mass and thus buoyancy transfers. In the case of an imposed buoyancy (salt) flux, the initial state has an interior buoyancy $b_1$ and a total buoyancy flux input $Q_1 =F_1 WL/2$ through the membrane with the denser (saltier) fluid located above the tank. In the initial steady state, the flux withdrawn is equal to the input $Q_1$.
In the case of a flux or a Neumann boundary condition, the buoyancy input is increased at time $t = 0$ from $Q_1$ to $Q_2 = Q_1 + \delta Q$. Previous experiments and numerical solutions show that the boundary
buoyancy input in the equilibrium state is carried to the bottom of the tank by the end-wall plume (shown in Fig. \ref{DELIF} and in previous works \cite{mullarney2004convection,stewart2012role,gayen2013energetics,PassaggiaHSW17}). Thus, we can assume that in the transient flow the ‘unbalanced’ buoyancy input (in excess of that in the equilibrium state) will also be carried to the bottom of the domain by the plume, where it spreads laterally across the bottom of the tank before being displaced downward by the continuing plume transport. Hence, the interior buoyancy $b(t)$ proceeds to increase, leading to an increasing buoyancy difference $b - b_f$ across the boundary layer and an increasing rate of conductive buoyancy withdrawal, which we write as $Q = Q_1 + Q_0(t)$.
The rate of change of buoyancy $\tilde{b}$ averaged over the interior assuming $\lambda_b \ll H$ writes as the buoyancy imbalance
\begin{equation}
LWH\frac{\mbox{d} \tilde{b}}{\mbox{d} t} =  \delta Q - Q'(t),
\label{eq:imbalance_1}
\end{equation}
where $\tilde{b}(t)$ is the spatial average of the buoyancy over the domain. The imbalance vanishes at large times, when $Q'(t) \rightarrow \delta Q$. For quasi-steady conduction in the boundary layer over the buoyant half part of the top, the rate of buoyancy withdrawal is 
\begin{equation}
Q_{1}+Q^{\prime}(t) \approx \beta \kappa\left(\tilde{b}-b_{f}\right) L W / 2 \lambda_b,
\label{eq:imbalance2}
\end{equation}
where we allow for freshening (or decrease of salt/buoyancy) over the half of the domain $L/2$ and write the gradient at the boundary as
$\langle \mbox{d}b/\mbox{d}z \rangle_{z=H} \approx \beta {(\tilde{b}-b_f)}/{\lambda_b}$. 
%
%
In their original analysis, Griffith {\it et al.}\cite{griffiths2013horizontal} did not separate the kinetic BL $\lambda_u$ from the buoyancy (thermal in their case) BL $\lambda_b$ and defined $\beta$ such that 95 \% of the overall buoyancy
difference lies in the BL. The constant was evaluated from DNS for $\beta\approx 1.4$ at $\rm{Pr}\approx 5$. Since the flow is essentially laminar in the stably stratified layer and transitional in the statically unstable zone, the constant $\beta$ can be seen as the ratio between the kinetic BL and the buoyancy BL which separates for both low- and high-Prandtl numbers such that
\begin{equation}
  \beta \approx \frac{\lambda_u\langle \mbox{d}b/\mbox{d}z \rangle_{z=H}} {{(\tilde{b}-b_f)}} \approx c_4 \lambda_u {\rm Nu} \approx c_4 {\rm Re}^{-1/2} {\rm Nu} \approx c_4 \rm{Pr^{\alpha}} \equiv c_4 \rm{Sc^{\alpha}}.
\end{equation}
where $\alpha=1/2$ in Rossby's and Shishkina \& Wagner's regimes while $\alpha=4/10$ for the Hughes' {\bf{\it et al.}} regime. Applying the above to the results of Griffith {\it et al.}, one obtains $c_4\approx 0.74$.
At large time, the system approaches the final equilibrium state, in which $\tilde{b} = b_2$ and
\begin{equation}
Q_{1}+\delta Q \approx \beta \kappa\left(b_{2}-b_{f}\right) L W / 2 \lambda_b.
\end{equation}
Taking $\lambda_b$ as constant for small changes in boundary conditions and combining eqs. (\ref{eq:imbalance_1})-(\ref{eq:imbalance2}) gives the interior buoyancy
\begin{equation}
\tilde{b} \approx b_{1}+\delta b\left(1-\mathrm{e}^{-\beta \kappa t / 2 \lambda_b H}\right).
\end{equation}
which exponentially approaches a final equilibrium temperature $b_2$ , the magnitude of the resulting change being
\begin{equation}
\delta b=b_{2}-b_{1}=2 \lambda_b \delta Q /\left(\beta \kappa L W\right)=\lambda_b \delta F /\left(\beta \kappa\right).
\end{equation}
In normalized form, the deviation from the final equilibrium is
\begin{equation}
\mathcal{B}=\left(\tilde{b}-b_{2}\right) /\left(b_{1}-b_{2}\right) \approx \mathrm{e}^{-\beta \kappa t / 2 \lambda_b H}.
\end{equation}
The imposed flux condition causes the box to equilibrate to the new conditions on the exponential time scale $\tau_{F} \approx \beta \lambda_b H / \kappa$. The Rayleigh number scaling (\ref{eq:Raf_scaling}) justifies our assumption of constant $\lambda_b$ for modest changes in boundary conditions. It also
implies more rapid adjustment for larger $\mathrm{Ra}_f$ such that
\begin{equation}
\kappa \tau_{F} / H^{2} \approx(2 / \beta)(\lambda_b / H) \approx  \Gamma \left(2 c_{2} / c_4\right) \mathrm{R a}_{F}^{-1 / 6}{\rm Sc}^{-1/2}.
\end{equation}
The evolution of ${\rm Ra}_f$ in each well is shown in Fig.\ref{fig:Raf} where both the source and sink of buoyancy reach the same value which implies that the flow has reached a steady state and that no evaporation is taking place (which was corrected from early experimental trials). Note the time scale was not non-dimensionalized to reflect the duration of the experiments in different tanks. For instance in the small tank, the diffusion time scale for salt $\tau_d\sim \kappa^{-1}H^{2}\approx 5$ days while in the largest tank would be nearly one year. while four days are necessary to obtain a steady state in the small tank, already hinting that a viscous scaling will be at stake, it took a month and a half to obtain a steady in the larger tank which confirms that vigorous convection at the surface is present. 
Under the conditions of the experiments reported in this paper we find $\left.\kappa \tau_{F} / H^{2} \approx [1.1\, 0.17]\times 10^{-2}\text { (or } \tau_{F} \approx [5.6 \times 10^{3}\, 9.5 \times 10^{4}] \mathrm{s}\right)$. Sample results are shown for the $L=0.5$m tank in Fig. \ref{fig:Raf}(a) and for the $L=2$m tank in Fig. \ref{fig:Raf}(b) where all cases where run for at least $100$ times the flux timescale $\tau_f$ than highlighted in the analysis. 
Note that $\tau_f$ was also used to estimate the time it would take between the moment particles where inserted and data collection for PIV was considered.

As recently suggested by Rocha {\it et al.} \cite{rocha2019heat}, this time scale may prove to be short compared to the actual time necessary to establish a complete stead state since the bulk and the boundary layers may be characterized by different time constants. In their numerical simulations, they found that a complete steady state was achieved for a time scale $\tau\approx 0.15\tau_d$. Note that these experiments were in a transitional regime at $\rm{Pr}=1$ and that our experiments were run for at least $0.15\tau_d$ until steady state analysis were carried out and confirming that steady states were reached for each experiment.

%

\subsection{Steady states and local measurements}\label{sec:steady_states}
\begin{figure*}[t!]

\includegraphics[width=\textwidth]{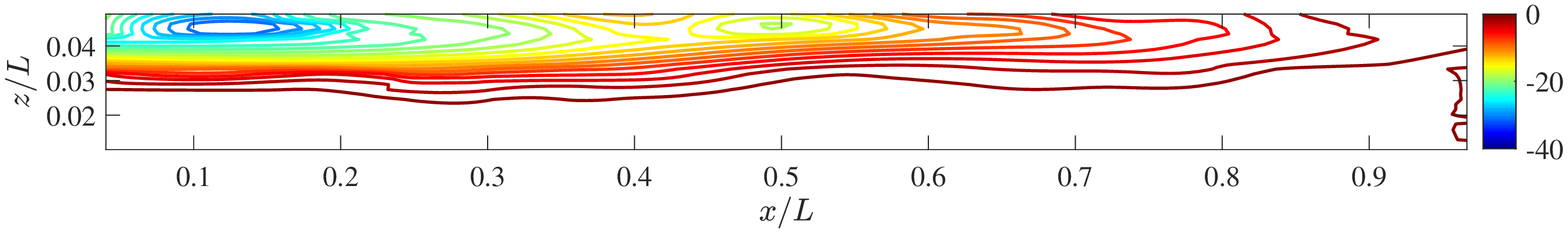}%
\put(-70,5){$\Psi/\kappa Ra^{-1/5}$}\put(-460,0){(a)}\\
\includegraphics[width=\textwidth]{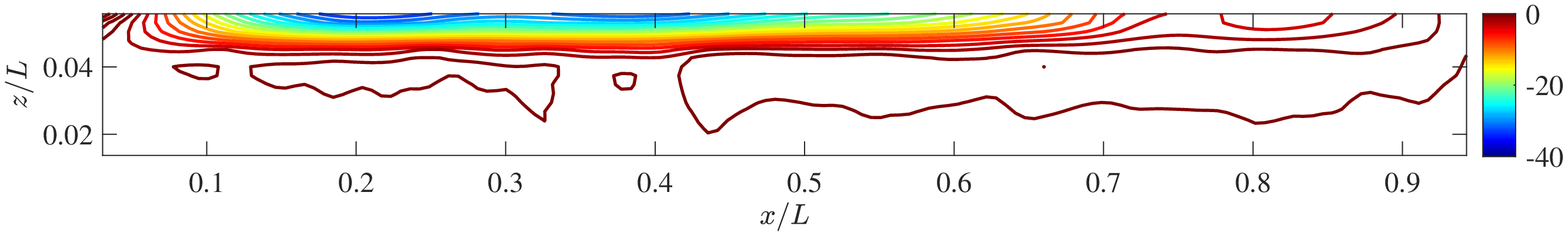}%
\put(-70,5){$\Psi/\kappa Ra^{-1/5}$}\put(-460,0){(b)}
\caption{Snapshots of the mean stream function for (a) ${\rm Ra}=1\times 10^{13}$ (b) ${\rm Ra}=3\times 10^{15}$ showing the transition regime between the plume dominated regime to the background thermal transport regime.}
\label{3D}
\end{figure*}

PIV of the full domain could only be performed for the smaller tank and we resort to another alternative to estimate the Reynolds and P\'eclet numbers for the larger tanks. Since we are working in high aspect ratio tanks (i.e. $A=16$), we propose an estimate for the Reynolds number which can be estimated from a measurements of the local stream function, taken far from the end wall at $x=\pm L/2$. At first order, the Reynolds number is approximated as 
\begin{equation}
{\rm Re}\approx \frac L V \int_0^H\int_{0}^{W}\int_{0}^{L}  \frac{x}{\nu} \frac{\partial \Psi}{\partial z}\Big{|}_{x=L/2,y=W/2}\mbox{d}V \approx \frac{L^2}{2H\nu} \int_0^H\mbox{d}\Psi\Big{|}_{x=L/2,y=W/2} \approx \frac{L^2}{H\lambda_u\nu}\max(\Psi(z))\Big{|}_{x=L/2,y=W/2}
\end{equation}
where the lateral effects were neglected.
The P\'eclet number can be defined equivalently such that $\rm Pe = Pr Re \equiv Sc Re$.
This approximation seems consistent with our experimental observation of the streamfunction that can seen from the PIV in Fig. \ref{3D}(a), at least for of the lowest $\rm Ra$.

\begin{figure*}[t!]
\scalebox{0.7}{\Large\input{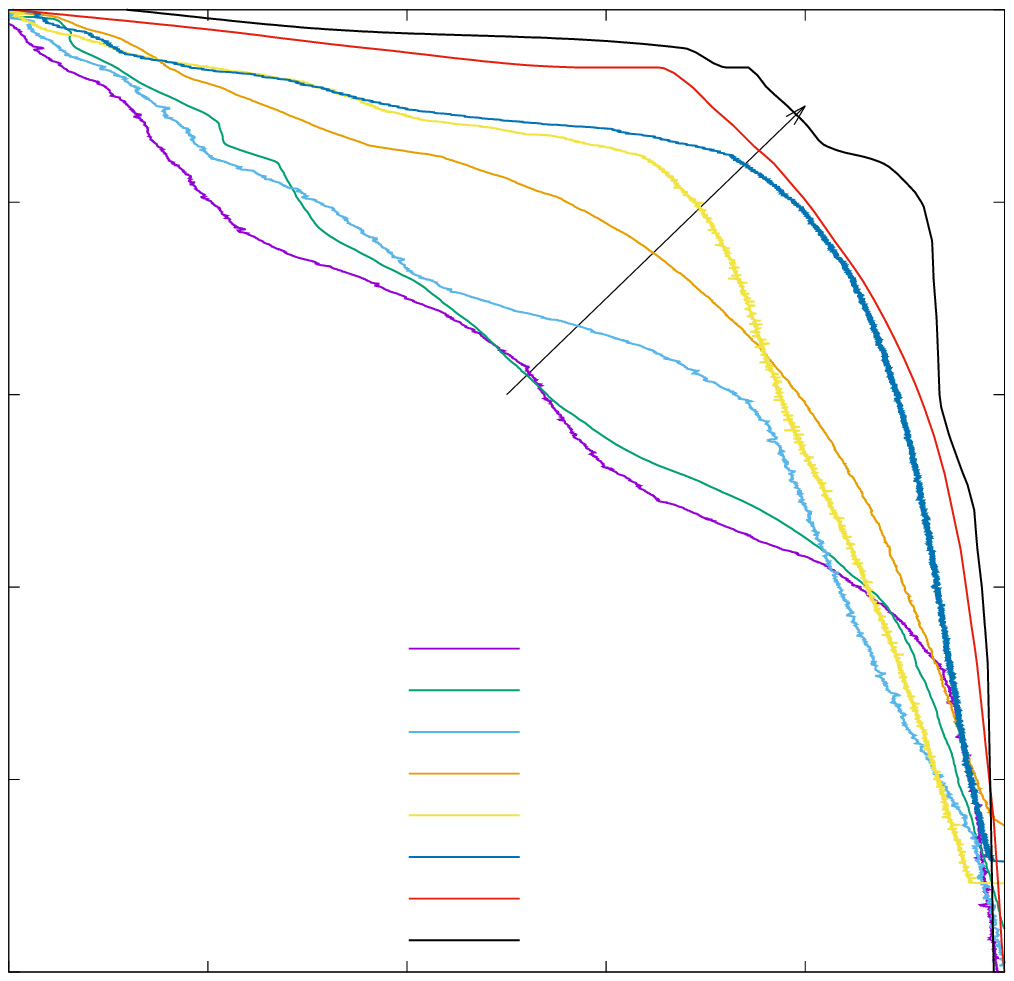}}\scalebox{0.7}{\Large\input{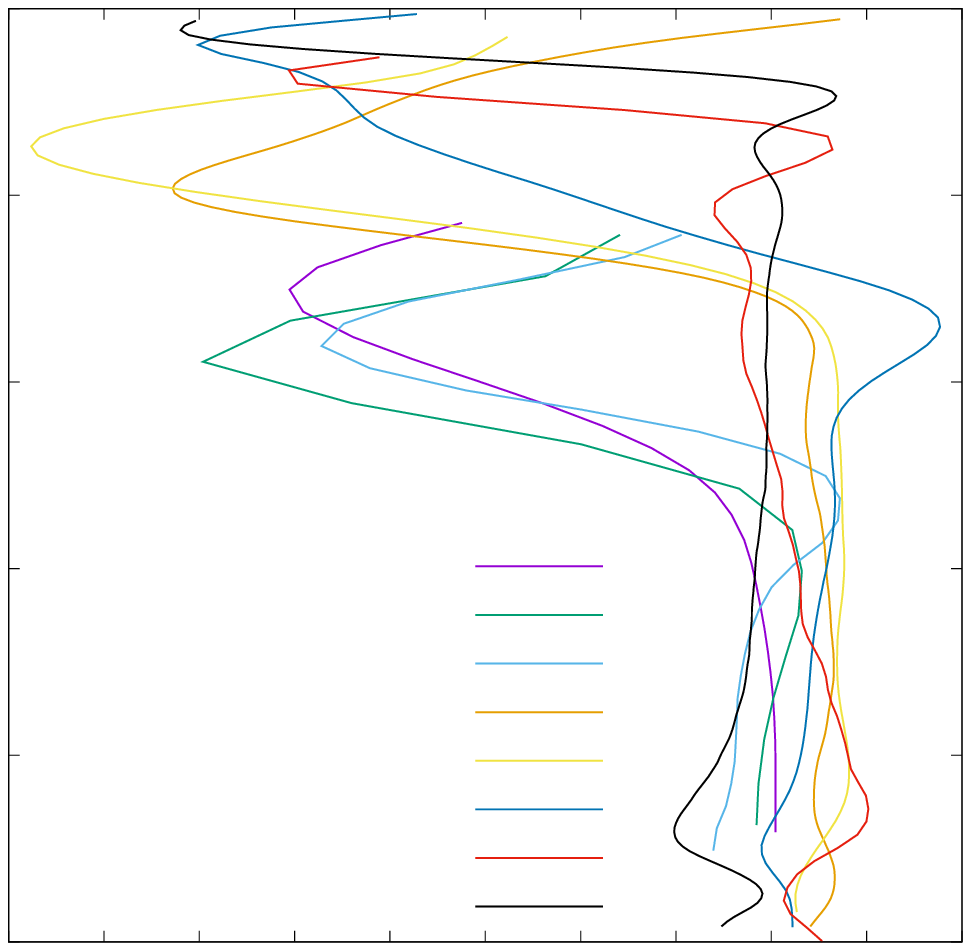}}
\put(-510,10){(a)}\put(-255,10){(b)}\\
\caption{(a) Normalized $({\rho(z)-\rho_{min}})/({\rho_{max}-\rho_{min}})$ density profiles and (b) streamfunction $\Psi(z)$ computed from averaged PIV data and rescaled with the Rossby scaling  as suggested by Chiu-Webster collected at the middle of the tank (i.e. at $x=L/2$) for most of the range collected in this study.} 
\label{b_Psi}
\end{figure*}

Rescaled density profiles, measured at the center of the domain $x=L/2$ are shown in Fig. \ref{b_Psi}(a) for most of the range of Rayleigh numbers reported in the present study. As $\rm Ra$ increases, the flow exhibits the same behaviour as seen in the experiment of Mularney {\it et al.}\cite{mullarney2004convection} but with a thinning of the pycnocline and an increase of the dense, well-mixed fluid, filling the bottom and the center of the domain.

Rescaled values of the streamfunction profiles with the Rossby scaling and collected at the same location are shown in Fig. \ref{b_Psi}(b) for the same values of $\rm Ra$ as in Fig. \ref{b_Psi}(a). 
The two regimes identified with the analysis of Nusselt number scaling can also be observed with the evolution of the streamfunction with respect to $\rm Ra$. The maximum of the streamfunction becomes progressively closer to the forced boundary at $z=H$ or $zA=1$. It is worth noting that the location of the streamfunction approaches the buoyancy-forced boundary when ${\rm Ra}_f\gtrsim 10^{17}$ or  ${\rm Ra}\gtrsim 10^{15}$ while the flow is essentially at rest in the core of the domain. This observation follows the results from the experiment of Wang \& Huang \cite{wang2005experimental} which were performed at $\rm Pr\approx 8$. Note that Wang \& Huang also reported a regime transition but did not see a change of exponent across each regime. We emphasize that in the case of Wang \& Huang, the aspect ratio was small $A=1.3$ and a similar width $W/L\approx1/8$ whereas in our experiments uses $A\approx16$ and $W/L\approx1/16$ and is therefore subject to the confinement effect as recently described in Shishkina \& Wagner\cite{ShishkinaW16}.
Note that there is a small recirculation region observed in \ref{b_Psi} at the bottom of the tank (shown by a small small bump in the streamfunction), which was present in nearly all experiments except for the smaller tank. We hypothesize that this feature, not present in the Wang \& Huang experiments, is possibly due to an undesirable heating from the bottom of the tank. 

\subsection{Scaling analysis}\label{sec:scaling}

\subsubsection{Flux-Rayleigh number analysis}\label{sec:scaling_Raf}

\begin{figure*}[t!]
\scalebox{0.7}{\Large\input{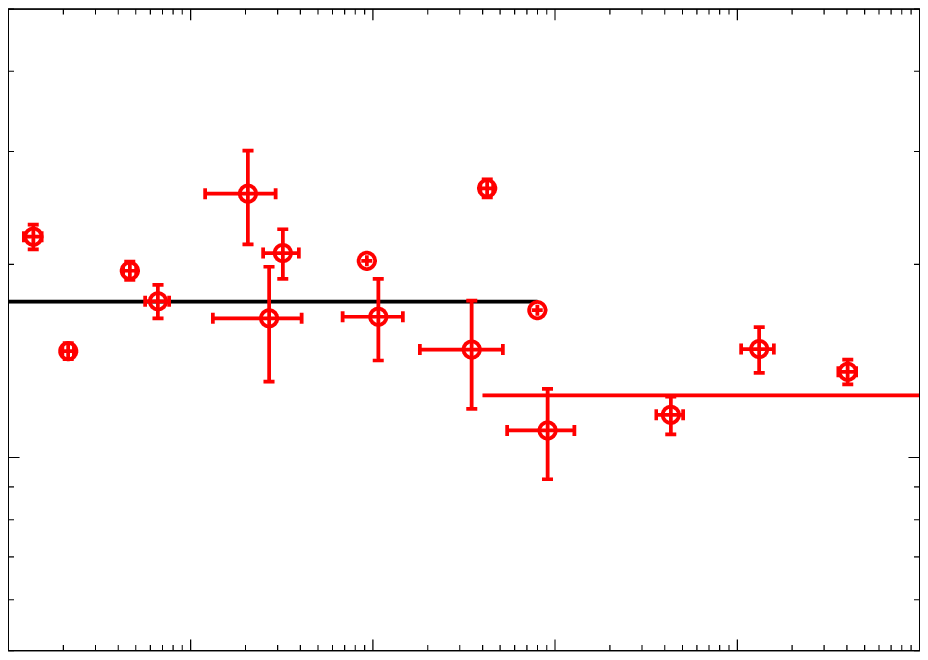}}\scalebox{0.7}{\Large\input{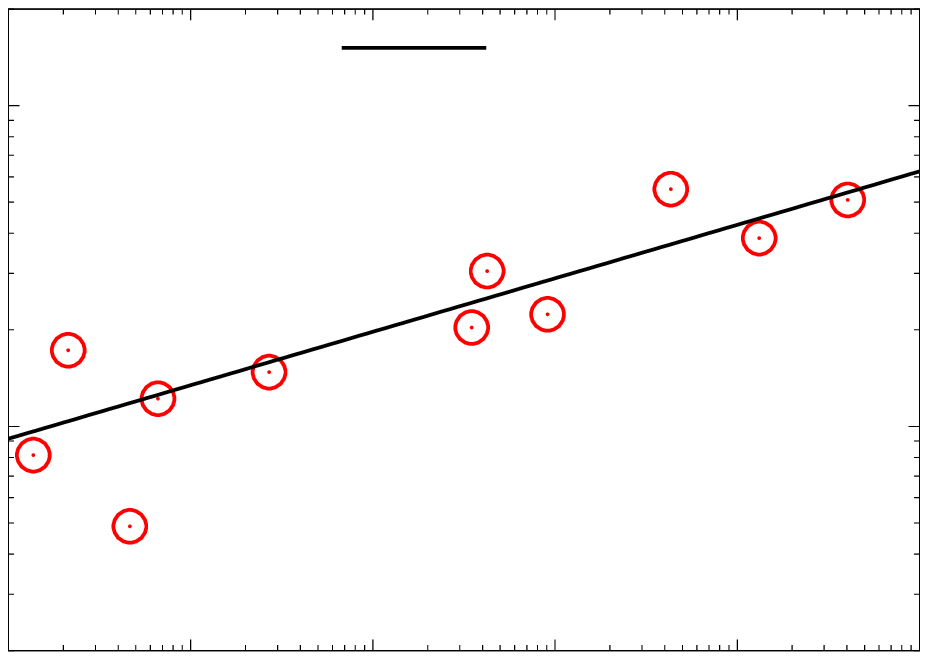}}
\put(-500,10){(a)}\put(-250,10){(b)}\\
\scalebox{0.7}{\Large\input{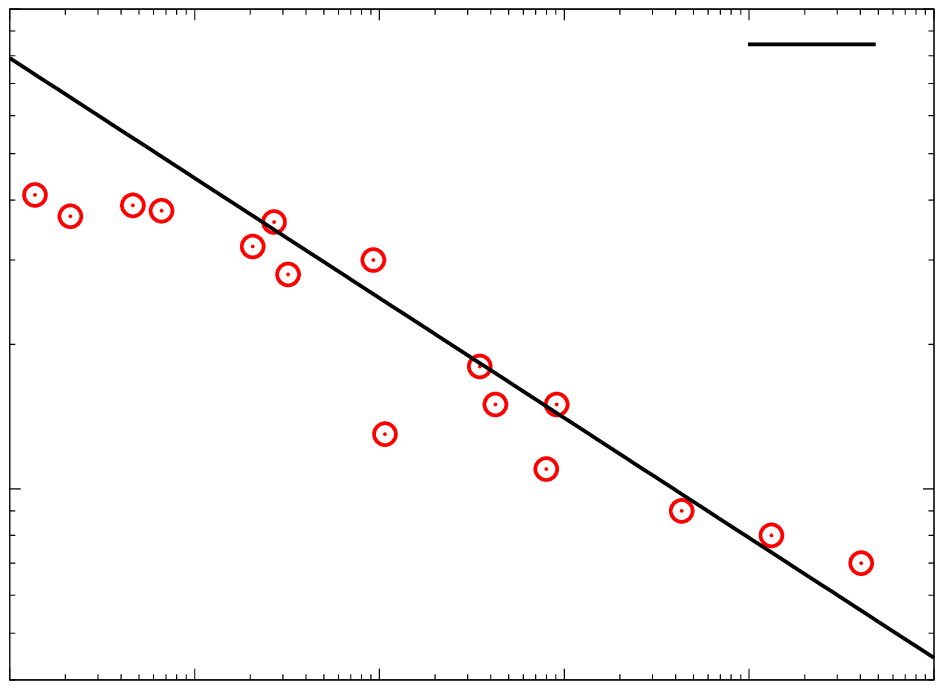}}\scalebox{0.7}{\Large\input{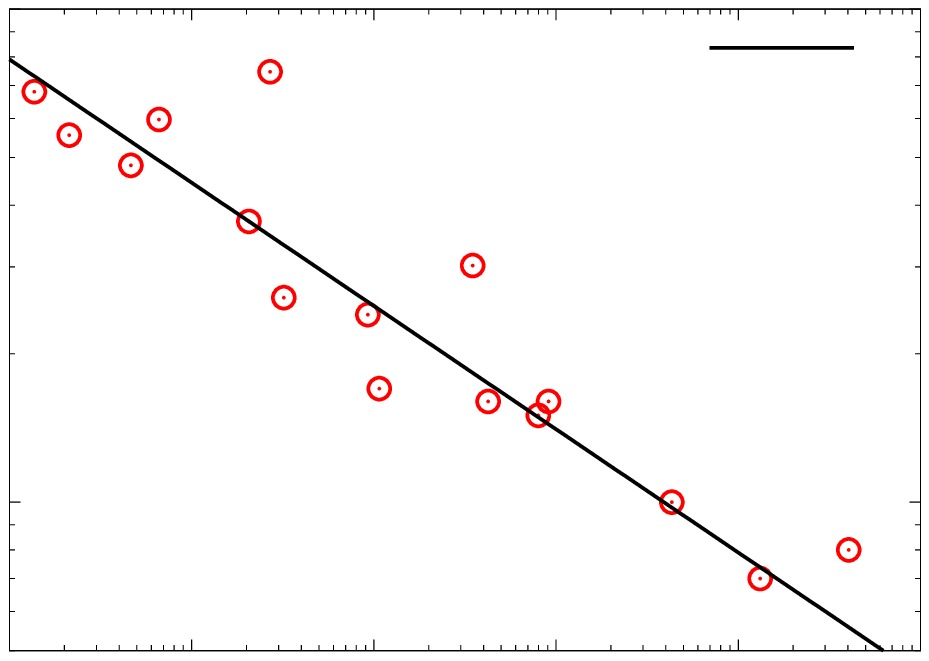}}
\put(-500,10){(c)}\put(-250,10){(d)}\\
\caption{(a) Evolution of the compensated ${\rm Nu}$ with the laminar scaling ${\rm Ra}_f^{-1/6}$ as a function of ${\rm Ra}_f$. (b) Streamfunction $\Psi_{\max}$ as a function of ${\rm Ra}_f$. (c) Buoyancy boundary-layer thickness $\lambda_b$ and (d) kinetic boundary-thickness $\lambda_u$, compared with the laminar scaling ${\rm Ra}_f^{-1/4}$ of Hughes {\it et al.}\cite{hughes2008horizontal}.
}
\label{Nu_Raf}
\end{figure*}

The evolution of the main flow quantities with respect to the flux Rayleigh number ${\rm Ra}_f$ are first analyzed motivated by the idea that the buoyancy flux is imposed across the membranes. Fig. \ref{fig:Raf}(a) shows a compensated plot of the evolution of the Nusselt for all experiments where a scaling law of the form ${\rm Nu}\approx c_2 {\rm Ra}_f^{-1/6}$ for nearly 5 orders of magnitude is found. This scaling is consistent with the scaling reported in Hughes {\it et al.} \cite{Hughes07} for a plume-type model where the laminar buoyancy boundary layer thickness sets the buoyancy and the kinetic energy exchanges.
We observe that the constant $c_2$ decreases abruptly from $1.9$ to $1.2$ for ${\rm Ra}_f> 10^{16}$ which may indicate a possible bifurcation of the flow from a regime to another.

The maximum of the streamfunction $\Psi_{max}$, measured at the center of the tank using the PIV and follows the same scaling \ref{fig:Raf}(b) such that $\Psi_{max} \approx c_1 {\rm Ra}_f^{1/6}$ where $c_1 \approx 0.07$. 
Note that this scaling is different than the ${\rm Ra}_f^{1/4}$ scaling reported in \cite{Hughes07}, and hints to a different regime. This could be already expected from the PIV measurements shown in Fig. \ref{3D}(a,b) where the plume only reaches half the depth of tank which is clearly different from the original experiments of Mullarney\cite{mullarney2004convection} performed at ${\rm Pr}\approx 5$ but is analog to the experimental work of Wang \& Huang \cite{wang2005experimental} ${\rm Pr}\approx 8$ and the thorough theoretical analysis of Chiu-Webster {\it et al.} \cite{chiu2008very} for ${\rm Pr}\rightarrow \infty$. The latter being naturally closer to the present study. Data will be further compared with this theory in the subsequent subsection.

The same observation follows for the thickness of the buoyancy boundary layer $\lambda_b$ shown in Fig. \ref{fig:Raf}(d) where the scaling is also seen to jump from $\lambda_b\sim{\rm Ra}_f^{0}$ for ${\rm Ra}_f< 10^{16}$ to $\lambda_b\sim{\rm Ra}_f^{-1/4}$ for ${\rm Ra}_f> 10^{16}$ which marks the separation from a full-depth circulation to a shallower type flow where the circulation progressively clusters underneath the surface. This feature was not not in the scope of the theoretical work of  Hughes {\it et al.}\cite{Hughes07} but was taken into account in the work of Chiu-Webster {\it et al.} \cite{chiu2008very}. Finally, the thickness of the streamfunction is found to scale across nearly five orders of magnitude as $\lambda_u\sim{\rm Ra}_f^{-1/4}$. 

This result may seem surprising at first since the thickness of the circulation decreases faster than the increase of the streamfunction. But it remains consistent with our PIV observations were the recirculating flow becomes shallower as ${\rm Ra}_f$ increases. HC at high-Prandtl numbers no-longer sustains a deep circulation for sufficiently high ${\rm Ra}_f$ and the plume no-longer spans the full depth of the domain. A such conclusion was also reached by Chiu-Webster {\it et al.} \cite{chiu2008very} who modelled theoretically very viscous HC or HC at high Prandtl numbers. Their theory was also backed by numerical simulations of the same flow and demonstrated similar features.
We therefore reconsider our analysis using their framework in the remainder of the paper and show that the flow does follow the physical characteristics of very viscous HC.

Before proceeding, the relation between the Rayleigh and the flux Rayleigh number can be tied to the Nusselt number using eq. \ref{eq:NuRafRa}.
In what follows, we reanalyse the above data as a function of the Rayleigh number, with the density measured in the wells. Revisiting the data allows for identifying new details, the characteristics of the flow, and thus the regimes previously found in the literature.

\subsubsection{Rayleigh number analysis}\label{sec:scaling_Ra}

\begin{figure*}[t!]
\scalebox{0.7}{\Large\input{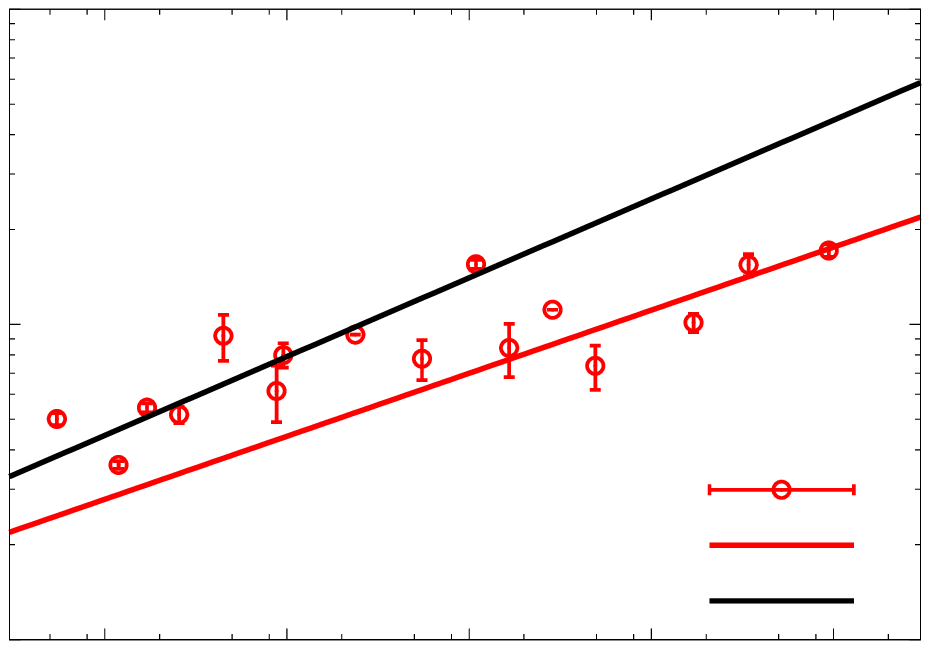}}\scalebox{0.7}{\Large\input{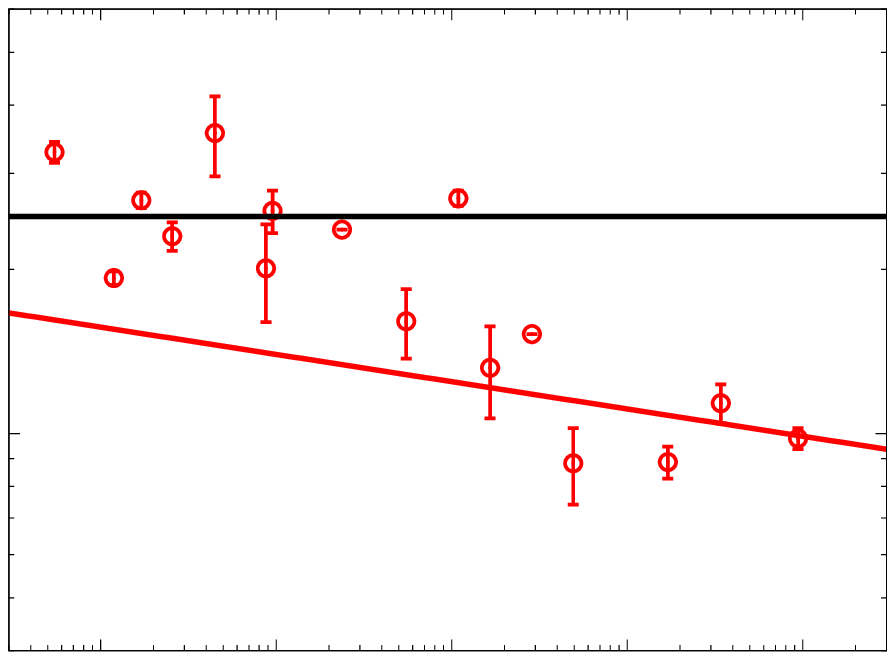}}
\put(-500,10){(a)}\put(-250,10){(b)}\\
\scalebox{0.7}{\Large\input{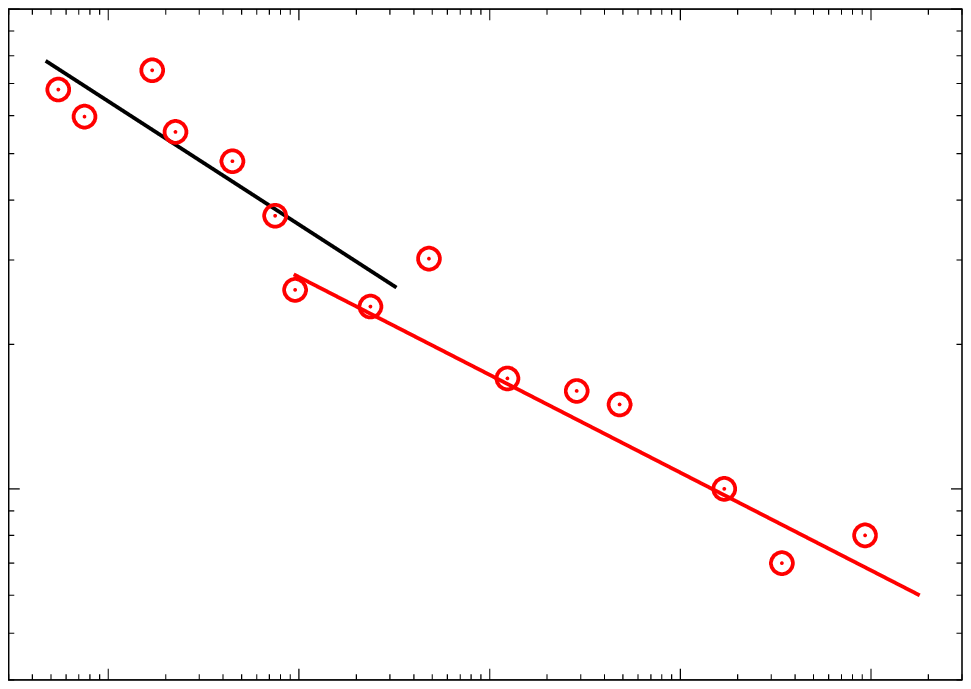}}\scalebox{0.7}{\Large\input{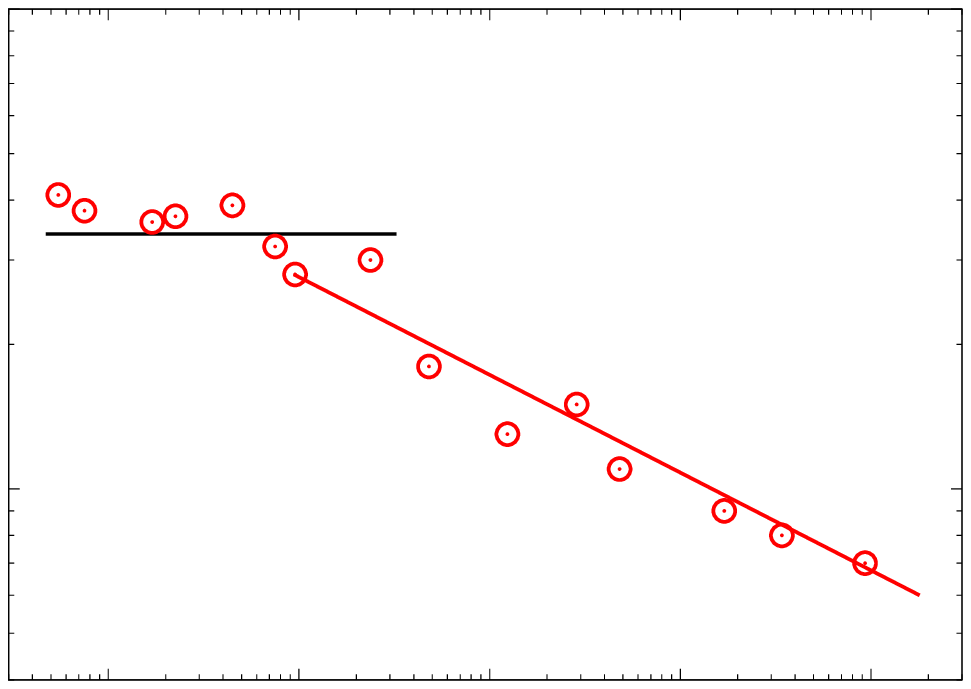}}
\put(-500,10){(c)}\put(-250,10){(d)}\\
\scalebox{0.7}{\Large\input{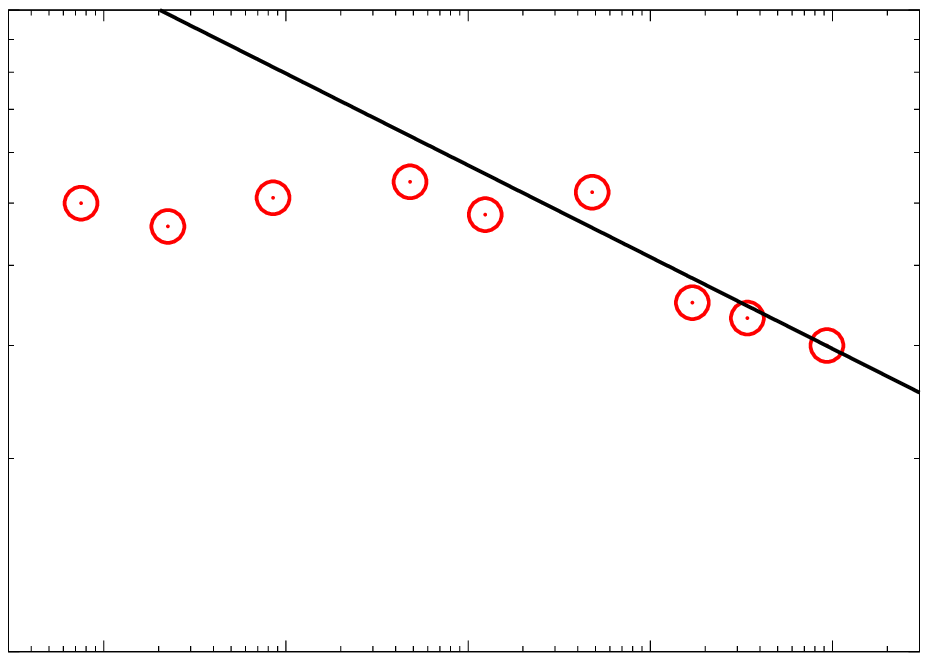}}\scalebox{0.7}{\Large\input{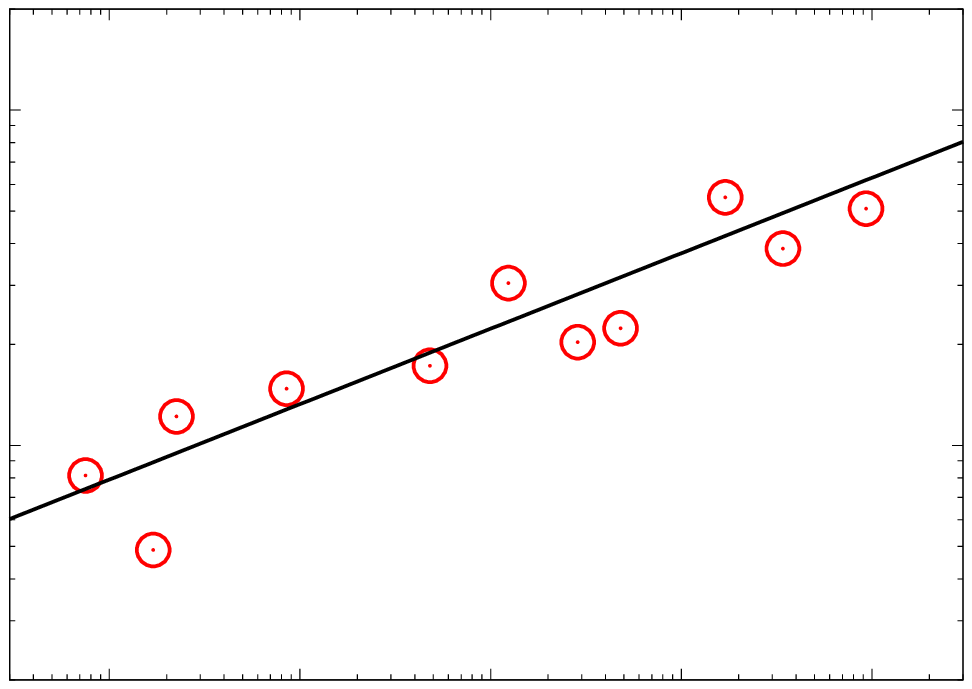}}
\put(-500,10){(e)}\put(-250,10){(f)}\\
\caption{(a) $\rm{Ra}$ and (b) rescaled-$\rm{Ra}$ dependencies of the Nusselt number. (c) Buoyancy boundary-layer thickness $\lambda_b$ and (d) kinetic boundary-thickness $\lambda_u$, compared with the theory of Chiu-Webster {\it et al.}\cite{chiu2008very} (red) an Shishkina \& Wagner\cite{ShishkinaW16} (black). (e) Bulk thickness $A(H-h)$ compared with the theory of Chiu-Webster {\it et al.}. (f)  Maximum of the stream function $\Psi_{\max}$ as a function of $\rm{Ra}$ and the linear regression over the full range. 
}
\label{Nu_Ra}
\end{figure*}

The dependence of $\rm{Nu}$ and $\Psi_{max}$ (and equivalently $\rm{Re}$) with respect to the Rayleigh number $\rm{Ra}$ are summarized in figure \ref{Nu_Ra}(a-d). The Nusselt number obtained in the experiment lies in the range ${\rm Ra}\approx[5.\,10^{12}, 8.\,10^{16}]$. Similarly to the Flux-Rayleigh number analysis, we observe two different scaling exponents (Fig. \ref{Nu_Ra}(a)): For $\rm Ra\lesssim 10^{15}$, the flow exhibits a scaling exponent ${\rm Nu}\sim {\rm Ra}^{1/4}$, while for $\rm Ra\gtrsim 10^{15}$ the rate of salt-uptake decreases to exhibit a $\rm Nu\sim Ra^{1/5}$-type scaling.
The same compensated plot is shown in Fig.\ref{Nu_Ra}(b) where the difference between the two scaling is only a factor $2.5$ at $\rm Ra=8. 10^{16}$ over the full dynamic range of our experiments and underlines the importance of the large Rayleigh-number ranges to differentiate between the two scaling laws.

Based on Figs.\ref{b_Psi}(a-b), we extrapolate the thickness of both the kinetic boundary layer $\lambda_u$ and the pycnocline thickness $\lambda_b$. The former is given by the location of the maximum of the streamfunction $\max{(\Psi(z))}$, the latter was extrapolated using the height of the fiftieth percentile of the rescaled density profile. The evolution of the buoyancy boundary layer is reported in Fig.\ref{Nu_Ra}(c) and shows again two different behaviours. For $\rm Ra<10^{15}$, the buoyancy boundary layer thickness decreases and follows a $\rm Ra^{-1/4}$ type scaling. At $\rm Ra \approx 10^{15}$, we observe a regime shift where the thickness now decreases such that $\rm Ra^{-1/4}$. A similar behaviour is observed in Fig.\ref{Nu_Ra}(d) where the thickness of the kinetic boundary layer $\lambda_u$ is reported. For $\rm Ra\lesssim 10^{15}$ the kinetic boundary layer saturates at $\lambda\approx0.4$, which is agreement with the study of Shishkina \& Wagner\cite{ShishkinaW16}. Past  $\rm Ra\gtrsim 2\; 10^{14}$, we recover the Rossby scaling for $\lambda_u\sim Ra^{-1/5}$ and together with the bulk reduction, suggests that we are observing the intrusion-type flow studied by Chiu-Webster {\it et al.}\cite{chiu2008very}.

We confirm this assertion in Fig. \ref{Nu_Ra}(e) where the thickness of the core $H-h$ is shown. This thickness was measured as the height of streamfunction shown in Fig. \ref{Nu_Ra} relating to the absolute value of the streamfunction $|\Psi|$ becoming smaller than $5\%$ of the maximum of the streamfunction. Using their plume theory, Chiu-Webster {\it et al.}\cite{chiu2008very} showed that the thickness of the bulk decreases as $\rm Ra^{-1/7}$. This scaling is compared with the few data points that suggested  such a behaviour for $\rm Ra\gtrsim 10^{15}$. Although only five data points seem to agree, these evidence suggest that we may be witnessing the intrusion regime described by Chiu-Webster\cite{chiu2008very}.

Conclusions from the maximum of the streamfunction shown in Fig. \ref{Nu_Ra}(f) are more difficult to draw. From Shishkina \& Wagner {\it et al.}\cite{ShishkinaW16}, we would expect a Reynolds number scaling as $\rm Re\sim Ra^{1/2}$ and $\lambda_u\sim {\rm Ra}^{1/2}$, resulting in $\rm \Psi_{max}\sim{\rm Ra}^{1/4}$. For larger values of $\rm Ra$, we would expect a $\rm{\Psi}_{\max} \sim \rm Ra^{1/5}$ similar to Rossby's scaling. A linear regression over the entire data set provides $\max{\Psi}\sim \rm Ra^{0.225}$, which is in between the $1/4$ exponent expected from Shishkina \& Wagner\cite{ShishkinaW16} analysis and the $1/5$ predicted in Chiu-Webster {\it et al.}\cite{chiu2008very}.\\

The present experimental results, based on a similar regime to Shishkina \& Wagner\cite{ShishkinaW16}, confirm a transition from the $I_l^*$ regime to the $I_u$ intrusion regime of Chiu-Webster regime\cite{chiu2008very} for Prandtl numbers one order of magnitude large than the direct numerical simulations of Shishkina \& Wagner\cite{ShishkinaW16} and Rossby\cite{Rossby65}, and Rayleigh numbers up to seven orders of magnitude larger than previous experiments and direct numerical simulations for similar Prandtl numbers. In the next subsection, we frame these results in the broader picture of horizontal convection and propose an updated regime transition diagram of horizontal convection.

\section{Completing the regime diagram of natural horizontal convection}

In this section, we summarize the different  HC regimes that were observed in the 
$({\rm Ra, Pr})$ space from the present work and its companion paper\cite{Passaggia2019LimitigA}, together with all the results that we could gather from the literature on HC. This section aims at extending the regime diagram in \cite{Hughes07} and the limiting regimes provided in \cite{shishkina2017scaling}, analog to results provided in \cite{GL00}.\\

We first provide a nomenclature for the different regimes:
\begin{itemize}
 \item \framebox[6mm]{$0$} Nearly conducting (i.e. ${\rm Nu}\sim Cst$),
 \item \framebox[6mm]{$I$} Laminar,
 \item \framebox[6mm]{$II$} Transitional (Only turbulent in the plume),
 \item \framebox[6mm]{$III$} Enhanced Transitional (Only turbulent in the BL but never observed),
 \item \framebox[6mm]{$IV$} Both boundary layers and plume are turbulent.\\
\end{itemize}

Additional subscripts and superscript are identified as:
\begin{itemize}
 \item \framebox[6mm]{$_u$} upper as large Prandlt numbers,
 \item \framebox[6mm]{$_l$} lower as low Prandlt numbers,
 \item \framebox[6mm]{$^*$} When essentially controlled by the aspect ratio.
\end{itemize}

We begin with the conducting regime \framebox[6mm]{$0$}, which was analyzed in Siggers {\it et al.}\cite{siggers2004bounds}, Chiu-Webster {\it et al.}\cite{chiu2008very}, and more recently by Sheard \& King\cite{sheard2011horizontal} where $\rm Nu=1$. The onset is independent of the Prandtl number, the transition is smooth and weakly affected by the aspect ratio $A$, and occurs when  $\rm Ra\approx 10^{3}$. The conducting regime neighbors the \framebox[6mm]{$I^*_{l}$} regime which follows the scaling in eq. (\ref{NuRe1/4}a) or the \framebox[6mm]{$I_{l}$} regime in eq. (\ref{lam_scal}a) depending on the aspect ratio. The transition was found at constant $\rm Ra$ which is consistent with the results reported in the literature\cite{sheard2011horizontal,siggers2004bounds}. \\ 

\begin{figure}[t!]
\centering
\hspace{0mm}\includegraphics[width=0.6\textwidth]{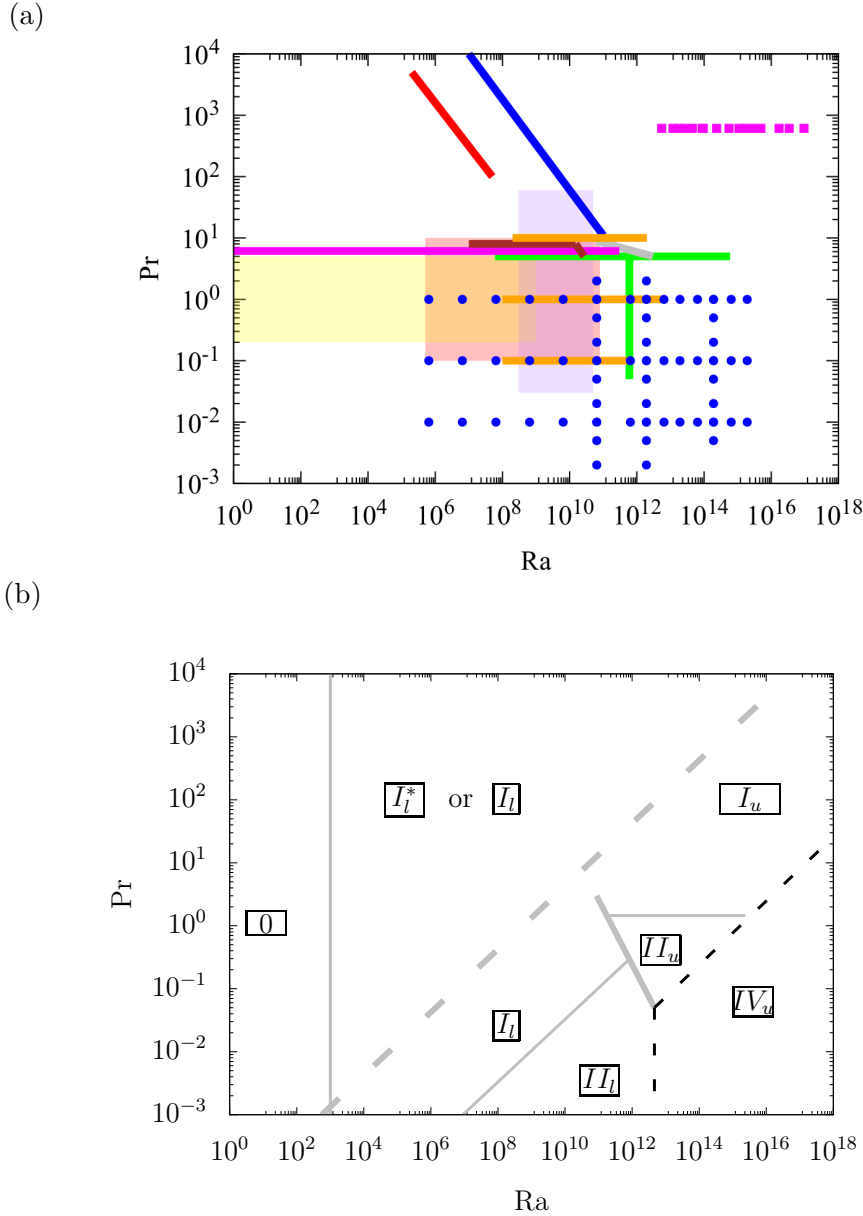}\put(-335,210){(a)}\\
\hspace{-10mm}\scalebox{0.9}{\large\input{Ra_Pr_landscape.tex}}
\put(-340,240){(b)}%
\caption{(a) Summary combining of all calculations found in the literature from experiments and DNS on horizontal convection together with the present experiments (\textcolor{magenta}{$\blacksquare$}) and DNS from the companion paper\cite{Passaggia2019LimitigA} (\textcolor{blue}{$\newmoon$}). These results also gather: (\textcolor{blue}{---}) (blue line) Rossby\cite{Rossby65} (1965) experiments, (\textcolor{red}{---}) (red line) Miller\cite{miller1968thermally} (1968), Beardsley \& Festa\cite{beardsley1972numerical} (1972) (\textcolor{black}{---}) (black lines), Paparella \& Young \cite{PaparellaY02} (2002) (red transparent square), Sigger {\it et al.} \cite{siggers2004bounds} (transparent yellow square), Mullarney {\it et al.} \cite{mullarney2004convection} (2004) (grey line), Wang \& Huang \cite{wang2005experimental} (2005) (brown line), Sheard \& King \cite{sheard2011horizontal} (2011) (magenta horizontal line), Gayen {\it et al.}\cite{Gayen14} (2014) (green lines),
Shishkina \& Wagner\cite{shishkina2017scaling} (2016) (transparent purple square), and Reiter \& Shishkina \cite{reiter2020classical} (2020) (orange horizontal lines).
(b) Phase diagrams in the $(\rm{Ra}, \rm{Pr})$ plane combining the present results with the Part 1 companion paperer \cite{Passaggia2019LimitigA} at low Prandtl numbers showing the different regimes of HC for the present flow geometry.}
\label{Ra_Pr_all}
\end{figure}
 
The connection between the laminar regimes \framebox[6mm]{$I^*_l$} and \framebox[6mm]{$I_l$} is determined by matching the Reynolds numbers in these neighboring regimes. From eqs. (\ref{NuRe1/4}b) and (\ref{lam_scal}b), we obtain the slope of the transition region between the regimes $I^*_{l}$ and the $I_l$, which is $\rm{Pr}\sim\rm{Ra}^{1/2}$. This transition is smooth\cite{shishkina2017scaling}, but strongly affected by both the aspect ratio of the domain and the type of boundary conditions. Note that free-slip boundary conditions are favorable to observe this regime. 
In fact, results from Rossby\cite{Rossby98}, Shishkina \& Wagner\cite{ShishkinaW16}, and the present work confirm that high aspect ratio domains are necessary to observe the  $I^*_l$ regime. For increasing aspect ratios, the boundary layer becomes increasingly thick with respect to the depth $H$ and eventually spans the full depth of the domain for small-enough Rayleigh number. To reflect this dependence in the regime diagram, this transition is marked by a dashed line due to its strong aspect ratio dependence. The present dashed line was approximated as a fit from the present results and Shishkina {\it et al.} for aspect ratios $A\sim\mathcal{O}(10)$ but the reader should keep in mind that the present landscape should include a third dimension as A to be complete and we estimate the present diagram to be valid for $A\sim [0.2,\, 20]$.\\

As both $\rm{Ra}$ and $\rm Pr$ increase, the Rossby regime progressively evolves towards the intrusion regime studied by Chiu-Webster {\it et al.}\cite{chiu2008very} and observed for large Rayleigh numbers in this study. This regime has the same Prandlt and Reynolds number dependence than the laminar Rossby regime but the flow has a significantly different structure. We therefore propose to name this flow \framebox[6mm]{$I_u$} to remain consistent with the Shishkina {\it et al.}\cite{ShishkinaGL16} nomenclature.\\

As $\rm Ra$ increases and for $\rm Pr\approx\mathcal{O}(1)$, we observe the \framebox[6mm]{$II_u$} regime Hughes {\it et al.}\cite{Hughes07}. As shown in Gayen {\it et al.}\cite{Gayen14}, the transition for increasing $\rm Ra$ is not smooth and this flow exists for a small part of the regime diagram. The flow transitions to the $I_u$ when $\rm Pr$ increases and appears at constant $\rm Pr$. This can be shown by equating eqs. (\ref{lam_scal}b) and (\ref{NuRe1/5h}b) and this transition from the $II_u$ to the $I_u$ was highlighted by Gayen {\it et al.}\cite{Gayen14} (see pp. 712).\\

For low $\rm Pr$, a sharp transition from the $I_l$ to the \framebox[6mm]{$II_l$} regime was reported in the companion paper\cite{Passaggia2019LimitigA}, which is presently under investigation using stability and bifurcation analyses\cite{PassaggiaSW17}. This transition occurs at $\rm Pr\sim Ra^{1/2}$ by matching again the Nusselt numbers on each side of the regime transition. Note that similarly to the high $\rm Pr$ regimes, the circulation shrinks as $\rm Pr$ decreases. Also note that the transition from $II_l$ to $II_u$ occurs at $\rm Ra^{-1}$. See the companion paper\cite{Passaggia2019LimitigA} for more details. \\

The last transition appears as the Rayleigh number increases and was first observed at ${\rm Ra}\approx 10^{11}$ and ${\rm Pr}\approx 0.1$. Both the $I_u$ and $II_l$ transition to the limiting regime \framebox[7mm]{$IV_u$}. This transition was found to be smooth and marks the appearance of a first limiting regime for asymptotic $\rm Ra$ verifying the zeroth law of turbulence. The transition from $II_l$ to $IV_u$ appears as $\rm Pr \sim Ra^{-1/2}$, while the transition from $II_l$ to $IV_u$ occurs for $\rm Pr \sim Ra^{-1/3}$ and $II_u$ to $IV_u$ for $\rm Pr \sim Ra^{1/4}$ is shown by the dashed line. This transition is shown in the companion paper to be associated with the reduction of the bulk induced by the turbulent plume \cite{Passaggia2019LimitigA}. Finally for high $\rm Pr$, matching the Nusselt numbers between $I_u$ and $IV_u$ provides $\rm Pr \sim Ra^{1/2}$ and is also shown by the dashed line for the same reasons.\\

We establish an updated regime diagram, as originally proposed by Hughes \& Griffiths\cite{hughes2008horizontal} and updated by Shishkina {\it et al.}\cite{ShishkinaGL16} (see  Fig. \ref{Ra_Pr_all}). This updated regime diagram encompasses, to the best of our knowledge, all previous DNS and experimental studies performed on natural horizontal convection. The last regime observed at high Rayleigh numbers is a turbulent dominated regime in the bulk which verifies the zeroth-law of turbulence \cite{ShishkinaGL16}. The resulting flow characteristics  at high Rayleigh numbers and all Prandtl numbers are an intensified turbulent near-surface circulation which follows Sandstr\"om's original inference\cite{Sandstrom16}, Jeffrey's argument\cite{jeffreys1925fluid} and the Paparella \& Young theorem\cite{PaparellaY02}. They argued that the flow generated by horizontal convection would result in an essentially stagnant pool of water with little to no flow in the core and an intensified turbulent circulation localized near the differentially heated boundary. Our results show that limiting regimes at all Prandtl numbers for large Rayleigh numbers inevitable lead to a such scenario. 


\section{Conclusions}

The present work considers the experimental study of horizontal convection at high-Prandtl and high-Rayleigh numbers. The aim of this work is to explore a part of the regime diagram which has not been yet been explored in previous studies. Related experimental work at high Prandtl numbers dates back to the original work of Rossby\cite{Rossby65} (1965) and Miller\cite{miller1968thermally} (1968) more than fifty years ago. Instead of differential heating where increasing the Prandtl number is achieved by increasing the viscosity of the working fluid, we consider solutal convection and use permeable dialysis membrane to allow for a mass/salinity flux through the forcing boundary while ensuring a no-flow/no-slip boundary condition in a high-aspect ratio and narrow domain.\\

Experiments are performed spanning four orders of magnitude in the Rayleigh number while keeping the Prandtl number constant. This allows for measuring two known regimes, already identified in the literature in natural horizontal convection at higher Prandtl numbers but for three to seven orders of magnitude higher than previously theorized or simulated. We report experimental evidence of the laminar regime recently identified by Shishkina \& Wagner\cite{ShishkinaW16}  where the recirculating flow and thus the boundary layer are of the same order of magnitude than the dept of the domain. This regime leads to a $\rm Nu \sim Ra^{1/4}$ regime which is consistent with direct numerical simulations at high $\rm Pr$ and lower $\rm Ra$ where the effects of confinement enhance the amount of laminar dissipation and provides a mechanism for an enhanced heat-transfer scaling than previously theorized by Rossby\cite{Rossby65}(1965) for small aspect ratio domains. Increasing further the Rayleigh number beyond $\rm Ra \gtrsim 10^{15}$ and for Schmidt numbers $\rm Sc\approx 610$, the flow exhibits a transition back to the Rossby regime as the boundary becomes progressively thinner. Experimental evidence of this regime is made possible following the work of Chiu-Webster {\it et al.}\cite{chiu2008very} (2008) who analyzed horizontal convection in the limit of asymptotically large Prandtl numbers, also known as very viscous horizontal convection. This regime is also known as the intrusion regime\cite{hughes2008horizontal}, follows the same scaling laws than the original Rossby regime by the structure of the flow is substantially different with a narrow recirculating/intrusion regime while the core of the flow remains essentially at rest. Using PIV and conductivity measurements at the center of the domain, we show that for high-Rayleigh numbers, the flow follows similar scaling laws than derived in Chiu-Webster {\it et al.}'s asymptotic analysis for the behaviour of the boundary layer and the thickness of the fluid at rest contained in the bulk. Definite conclusions about the magnitude of the streamfunction are harder to draw due to measurements uncertainties but hint to the same arguments.\\

We therefore report a new regime transition in horizontal convection at Large Rayleigh and Prandtl numbers and use these results in combination with the companion paper where low Prandtl number regimes were investigated. We combine previous experimental and numerical evidence where regimes and their transitions were identified. In this scope, we propose an updated regime diagram compared with the previous regimes originally proposed by Hughes \& Griffiths\cite{hughes2008horizontal} and more recently by Shishkina {\it et al.} \cite{ShishkinaGL16}. Seven distinct regimes are mapped for which observations using either DNS or experiments have been confirmed and use their Nusselt number dependencies to draw a complete regime diagram from seven orders of magnitude for Prandtl numbers and seventeen orders of magnitude for the Rayleigh number. We also report the six limiting regimes known to this date and in particular, the transition to the turbulent limiting regime $IV_u$ for asymptotically large Rayleigh numbers and put an emphasis on the role of the aspect ratio of the domain for large Prandtl numbers. While it was also shown that the ultimate regime $IV_l$ cannot be achieved (see ref.\cite{siggers2004bounds,passaggia2016global,Passaggia2019LimitigA}), the present regime diagram is consistent with previous studies and reviews and provides a complete parametrisation of horizontal convection which can be used for engineering and geophysical applications.\\

The authors acknowledge the support of the National Science Foundation Grant Number OCE--1155558 and OCE--1736989.

\bibliographystyle{plain}

\bibliography{bib}

\providecommand{\noopsort}[1]{}\providecommand{\singleletter}[1]{#1}%
\begin{thebibliography}{10}

\bibitem{beardsley1972numerical}
R.~C. Beardsley and J.~F. Festa.
\newblock A numerical model of convection driven by a surface stress and
  non-uniform horizontal heating.
\newblock {\em J. Phys. Oceanogr.}, 2(4):444--455, 1972.

\bibitem{carminati2017conduino}
M.~Carminati and P.~Luzzatto-Fegiz.
\newblock Conduino: Affordable and high-resolution multichannel water
  conductivity sensor using micro {USB} connectors.
\newblock {\em Sens. Act. B: Chem.}, 251:1034--1041, 2017.

\bibitem{chiu2008very}
S.~Chiu-Webster, E.~J. Hinch, and J.~R. Lister.
\newblock Very viscous horizontal convection.
\newblock {\em J. Fluid Mech.}, 611:395--426, 2008.

\bibitem{defant1961physical}
A.~Defant.
\newblock {\em Physical oceanography}, volume~1.
\newblock Pergamon, 1961.

\bibitem{Gayen14}
B.~Gayen, R.~W. Griffiths, and G.~O. Hughes.
\newblock Stability transitions and turbulence in horizontal convection.
\newblock {\em J. Fluid Mech.}, 751:698--724, 7 2014.

\bibitem{gayen2013energetics}
B.~Gayen, R.~W. Griffiths, G.~O. Hughes, and J.~A. Saenz.
\newblock Energetics of horizontal convection.
\newblock {\em J. Fluid Mech.}, 716, 2013.

\bibitem{gramberg2007convection}
H.~J.~J. Gramberg, P.~D. Howell, and J.R. Ockendon.
\newblock Convection by a horizontal thermal gradient.
\newblock {\em J. Fluid Mech.}, 586:41--57, 2007.

\bibitem{griffiths2015turbulent}
R.~W. Griffiths and B.~Gayen.
\newblock Turbulent convection insights from small-scale thermal forcing with
  zero net heat flux at a horizontal boundary.
\newblock {\em Phys. Rev. Lett.}, 115(20):204301, 2015.

\bibitem{griffiths2013horizontal}
R.~W. Griffiths, G.~O. Hughes, and B.~Gayen.
\newblock Horizontal convection dynamics: insights from transient adjustment.
\newblock {\em J. Fluid Mech.}, 726:559--595, 2013.

\bibitem{GL00}
S.~Grossmann and D.~Lohse.
\newblock Scaling in thermal convection: a unifying theory.
\newblock {\em J. Fluid Mech.}, 407:27--56, 2000.

\bibitem{harned1954diffusion}
H.~S. Harned.
\newblock The diffusion coefficients of the alkali metal chlorides and
  potassium and silver nitrates in dilute aqueous solutions at 25$^o$c.
\newblock {\em Proc. Nat. Acad. Sci.}, 40(7):551, 1954.

\bibitem{hughes2008horizontal}
G.~O. Hughes and R.~W. Griffiths.
\newblock Horizontal convection.
\newblock {\em Annu. Rev. Fluid Mech.}, 40:185--208, 2008.

\bibitem{Hughes07}
G.~O. Hughes, R.~W. Griffiths, J.~C. Mullarney, and W.~H. Peterson.
\newblock A theoretical model for horizontal convection at high rayleigh
  number.
\newblock {\em J. Fluid Mech.}, 581:251--276, 2007.

\bibitem{ilicak2012simulations}
M.~Ilicak and G.~K. Vallis.
\newblock Simulations and scaling of horizontal convection.
\newblock {\em Tellus A}, 64(1):18377, 2012.

\bibitem{jeffreys1925fluid}
H.~Jeffreys.
\newblock On fluid motions produced by differences of temperature and humidity.
\newblock {\em Quart. J. Royal Meteo. Soc.}, 51(216):347--356, 1925.

\bibitem{krishnamurti2003double}
R.~Krishnamurti.
\newblock Double-diffusive transport in laboratory thermohaline staircases.
\newblock {\em J. Fluid Mech.}, 483:287--314, 2003.

\bibitem{Landau87}
L.~D. Landau and E.~M. Lifschitz.
\newblock {\em Statistische Physik}.
\newblock Akademie-Verlag, 1987.

\bibitem{matusik2019response}
K.~E. Matusik and S.~G. Llewellyn-Smith.
\newblock The response of surface buoyancy flux-driven convection to localized
  mechanical forcing.
\newblock {\em Exp. Fluids}, 60(5):79, 2019.

\bibitem{meunier2003analysis}
P.~Meunier and T.~Leweke.
\newblock Analysis and treatment of errors due to high velocity gradients in
  particle image velocimetry.
\newblock {\em Exp. fluids}, 35(5):408--421, 2003.

\bibitem{miller1968thermally}
R.~C. Miller.
\newblock {\em A thermally convecting fluid heated non-uniformly from below.}
\newblock PhD thesis, Massachusetts Institute of Technology, 1968.

\bibitem{mullarney2004convection}
J.~C. Mullarney, R.~W. Griffiths, and G.~O. Hughes.
\newblock Convection driven by differential heating at a horizontal boundary.
\newblock {\em J. Fluid Mech.}, 516:181--209, 2004.

\bibitem{PaparellaY02}
F.~Paparella and W.~R. Young.
\newblock {Horizontal convection is non-turbulent}.
\newblock {\em J. Fluid Mech.}, 466:205--214, 2002.

\bibitem{PassaggiaHSW17}
P.-Y. Passaggia, M.~W. Hurley, A.~Scotti, and B.~L. White.
\newblock Experiments on horizontal convection at high {R}ayleigh and {P}randtl
  numbers.
\newblock {\em Phys. Rev. Fluids}, page Accepted, 2017.

\bibitem{passaggia2012transverse}
P.-Y. Passaggia, T.~Leweke, and U.~Ehrenstein.
\newblock Transverse instability and low-frequency flapping in incompressible
  separated boundary layer flows: an experimental study.
\newblock {\em J. Fluid Mech.}, 703:363--373, 2012.

\bibitem{passaggia2016global}
P.-Y. Passaggia, A.~Scotti, and B.~White.
\newblock Global stability and flow transition in horizontal convection.
\newblock In {\em International Symposium on Stratified Flows}, volume~1, 2016.

\bibitem{PassaggiaSW17}
P.-Y. Passaggia, A.~Scotti, and B.~L. White.
\newblock Transition and turbulence in horizontal convection: linear stability
  analysis.
\newblock {\em J. Fluid Mech.}, 821:31--58, 2017.

\bibitem{Passaggia2019LimitigA}
P.-Y. Passaggia, A.~Scotti, and B.~L. White.
\newblock Limiting regimes of turbulent horizontal convection. part i:
  Intermediate- and low-prandtl numbers.
\newblock {\em Submitted to Phys. Rev. Fluids}, 2019.

\bibitem{ramme2019transition}
L.~Ramme and U.~Hansen.
\newblock Transition to time-dependent flow in highly viscous horizontal
  convection.
\newblock {\em Phys. Rev. Fluids}, 4(9):093501, 2019.

\bibitem{reiter2020classical}
P.~Reiter and O.~Shishkina.
\newblock Classical and symmetrical horizontal convection: detaching plumes and
  oscillations.
\newblock {\em J. Fluid Mech.}, 892, 2020.

\bibitem{rocha2019heat}
C.~Rocha, N.~C. Constantinou, S.~G.~Llewellyn Smith, and W.~R. Young.
\newblock The heat flux of horizontal convection: definition of the nusselt
  number.
\newblock {\em arXiv preprint arXiv:1912.05229}, 2019.

\bibitem{rosevear2017turbulent}
M.~G. Rosevear, B.~Gayen, and R.~W. Griffiths.
\newblock Turbulent horizontal convection under spatially periodic forcing: a
  regime governed by interior inertia.
\newblock {\em J. Fluid Mech.}, 831:491--523, 2017.

\bibitem{Rossby65}
H.~T. Rossby.
\newblock On thermal convection driven by non-uniform heating from below: an
  experimental study.
\newblock {\em Deep-Sea Res.}, 12:9--16, 2 1965.

\bibitem{Rossby98}
T.~Rossby.
\newblock {Numerical experiments with a fluid heated non-uniformly from below}.
\newblock {\em Tellus}, 50A:242--257, 1998.

\bibitem{Sandstrom08}
J.W. Sandstr{\"o}m.
\newblock {Dynamische versuche mit meerwasser}.
\newblock {\em Ann. Hydrogr. Marit. Meteorol.}, 36:6--23, 1908.

\bibitem{Sandstrom16}
J.W. Sandstr{\"o}m.
\newblock Meteorologische studien im schwedischen hochgebirge.
\newblock {\em G\"oteborgs Kungl. Vet. Handl.}, 17(4):48, 1916.

\bibitem{sheard2011horizontal}
G.~J. Sheard and M.~P. King.
\newblock Horizontal convection: effect of aspect ratio on rayleigh number
  scaling and stability.
\newblock {\em App. Math. Modell.}, 35(4):1647--1655, 2011.

\bibitem{shishkina2017scaling}
O.~Shishkina, M.~S. Emran, S.~Grossmann, and D.~Lohse.
\newblock Scaling relations in large-prandtl-number natural thermal convection.
\newblock {\em Phys. Rev. Fluids}, 2(10):103502, 2017.

\bibitem{ShishkinaGL16}
O.~Shishkina, S.~Grossman, and D.~Lohse.
\newblock Heat and momentum transport scalings in horizontal convection.
\newblock {\em Geophys. Res. Lett.}, 43(3):1219--1225, 2016.

\bibitem{ShishkinaW16}
O.~Shishkina and S.~Wagner.
\newblock Prandtl-number dependence of heat transport in laminar horizontal
  convection.
\newblock {\em Phys. Rev. Lett.}, 116(2):024302, 2016.

\bibitem{siggers2004bounds}
JH~Siggers, RR~Kerswell, and NJ~Balmforth.
\newblock Bounds on horizontal convection.
\newblock {\em Journal of Fluid Mechanics}, 517:55--70, 2004.

\bibitem{stewart2012role}
K.~D. Stewart, G.~O. Hughes, and R.~W. Griffiths.
\newblock The role of turbulent mixing in an overturning circulation maintained
  by surface buoyancy forcing.
\newblock {\em J. Phys. Oceanogr.}, 42(11):1907--1922, 2012.

\bibitem{sverdrup1942oceans}
H.~U. Sverdrup, M.~W. Johnson, R.~H. Fleming, et~al.
\newblock {\em The Oceans: Their physics, chemistry, and general biology},
  volume~7.
\newblock Prentice-Hall New York, 1942.

\bibitem{wang2005experimental}
W.~Wang and R.~X. Huang.
\newblock An experimental study on thermal circulation driven by horizontal
  differential heating.
\newblock {\em J. Fluid Mech.}, 540:49--73, 2005.

\bibitem{whitehead2008laboratory}
J.~A. Whitehead and W.~Wang.
\newblock A laboratory model of vertical ocean circulation driven by mixing.
\newblock {\em J. Phys. Oceanogr.}, 38(5):1091--1106, 2008.

\end{thebibliography}

\end{document}